\documentclass[conference]{IEEEtran}
\IEEEoverridecommandlockouts
\usepackage{cite}
\usepackage{amsmath,amssymb,amsfonts}
\usepackage{algorithmic}
\usepackage{graphicx}
\usepackage{textcomp}
\usepackage{xcolor}
\usepackage{tcolorbox}
\usepackage{balance}
\usepackage{pifont}
\def\BibTeX{{\rm B\kern-.05em{\sc i\kern-.025em b}\kern-.08em
    T\kern-.1667em\lower.7ex\hbox{E}\kern-.125emX}}

\usepackage{enumitem}
\usepackage{subcaption}
\usepackage{subcaption}
\graphicspath{ {./figures/} }
\usepackage{url}
\usepackage[linesnumbered,lined,boxed,commentsnumbered]{algorithm2e}
\usepackage{mathtools} % Make sure you have this in your preamble

\usepackage[colorinlistoftodos,prependcaption, textsize=small, textwidth=30]{todonotes}

\definecolor{mygreen}{RGB}{0,128,0}

\newcommand\blfootnote[1]{%
  \begingroup
  \renewcommand\thefootnote{}%
  \footnotetext{#1}%
  \addtocounter{footnote}{-1}%
  \endgroup
}

\begin{document}

\title{CoLSE: A Lightweight and Robust Hybrid Learned Model for Single-Table Cardinality Estimation using Joint CDF}

%\and
%\and

% % Tighten figure/table caption spacing and floats to save space
% \setlength\abovecaptionskip{0pt}              % No extra space ABOVE captions
% \setlength\belowcaptionskip{-2pt}              % Slightly reduce space BELOW captions (can pull content up)
% \setlength{\floatsep}{1pt plus 0.5pt minus 1pt} % Space BETWEEN floats (figures/tables)
% \setlength{\textfloatsep}{1pt plus 0.5pt minus 1pt} % Space BETWEEN floats and surrounding text
% \setlength{\intextsep}{1pt plus 0.5pt minus 1pt}    % Space around in-text (H) floats

\author{
\IEEEauthorblockN{Lankadinee Rathuwadu, Guanli Liu, Christopher Leckie,  Renata Borovica-Gajic}
\IEEEauthorblockA{lrathuwadu@student.unimelb.edu.au,  \{guanli.liu1, caleckie, renata.borovica\}@unimelb.edu.au}
\IEEEauthorblockA{\textit{School of Computing and Information Systems} \\
	\textit{University of Melbourne}}}
\maketitle

\blfootnote{This is the author's accepted manuscript for ICDE 2026.
The final version will appear in the proceedings.}

%\begin{abstract}
%Cardinality Estimation (CE), the problem of estimating the result size of queries, is an important problem in query optimization. Accurate cardinality estimates are essential for generating efficient query execution plans. In recent years, machine learning techniques have been introduced for accurate cardinality estimation and can be broadly categorized into query-driven and data-driven approaches. Data-driven approaches learn a joint distribution of data, while query-driven approaches construct a regression model from a query to its cardinality. Ideally, a cardinality estimation technique should strike an optimal balance among three key factors: accuracy, efficiency, and memory footprint. However, the current state-of-the-art models struggle to achieve this balance effectively. As a solution, we propose CoLSE, a hybrid learned approach for single table cardinality estimation. Our approach directly models the joint probability over queried intervals or ranges by developing an algorithm with the use of copula models. Finally, the developed algorithm is coupled with a lightweight neural network to correct for any estimation errors. Experimental results demonstrate that CoLSE achieves a well-balanced trade-off between accuracy, training time, inference time, and model size compared to the current state-of-the-art models.
%\end{abstract}
\begin{abstract}
Cardinality estimation (CE)—the task of predicting the result size of queries—is a critical component of query optimization. Accurate estimates are essential for generating efficient query execution plans. Recently, machine learning techniques have been applied to CE, broadly categorized into query-driven and data-driven approaches. Data-driven methods learn the joint distribution of data, while query-driven methods construct regression models that map query features to cardinalities. Ideally, a CE technique should strike a balance among three key factors: accuracy, efficiency, and memory footprint. However, existing state-of-the-art models often fail to achieve this balance. 

To address this, we propose \textbf{CoLSE}, a hybrid learned approach for single-table cardinality estimation. CoLSE directly models the joint probability over queried intervals using a novel algorithm based on copula theory and integrates a lightweight neural network to correct residual estimation errors. Experimental results show that CoLSE achieves a favorable trade-off among accuracy, training time, inference latency, and model size, outperforming existing state-of-the-art methods.
\end{abstract}

\vspace{-0.5em}
\section{Introduction}

Cardinality estimation (CE) is a fundamental yet challenging component of cost-based query optimizers. The optimizer relies on cardinality estimates to evaluate alternative execution plans for a given query and select the most efficient one. Formally, CE refers to the task of estimating the result size of queries with multiple predicates, based on data statistics and assumptions about data distributions, column correlations, and join relationships. A closely related problem is selectivity estimation, which computes the fraction of tuples that satisfy the query predicates. 
CE has been described as the “Achilles’ heel” of query optimization~\cite{lohman2014query}, as it is responsible for many of the optimizer’s performance issues. Inaccurate cardinality estimates can lead to poor plan choices, resulting in significant performance degradation~\cite{leis2015good, wang2021face}. Consequently, selectivity and cardinality estimation have been active areas of research for several decades, with numerous studies examining the effectiveness of various techniques.

%Past works related to cardinality estimation can be divided into three main categories: Synopsis-based methods, Sampling-based methods, and Learning-based methods~\cite{lan2021survey}. And the methodologies have been developed upon two branches; single table cardinality estimation and join cardinality estimation. Single-table cardinality estimation concentrates on predicting the number of records returned from a single table, while join cardinality estimation involves estimating the number of records resulting from the combination of two or more tables. Single-table CE is a fundamental and long standing problem in query optimization~\cite{wang2020we, dutt2019selectivity, yang2019deep}. These estimations largely influence access path decisions, which form the base of the execution plan.  Errors at this level propagate upward via the query plan, often accumulating and resulting in poor performance.  Previous research~\cite{lee2023analyzing} has shown that accurate single-table cardinality estimates can avoid costly operations such as full table scans and enable more efficient join orders. Furthermore, recent work in join cardinality estimation~\cite{wu2023factorjoin, kim2024asm, zhu2021glue} focuses on developing techniques that integrate single-table cardinality estimates. Therefore, we believe that examining single-table CE is essential; thus, this study concentrates on that aspect.

Existing CE techniques can be broadly categorized into \textit{synopsis-based}, \textit{sampling-based}, and \textit{learning-based} methods~\cite{lan2021survey}, each developed for either \textit{single-table} or \textit{join cardinality estimation}. 

Single-table CE aims to estimate the number of records returned from a single relation, given one or more local predicates. It plays a foundational role in query optimization, as these estimates influence early-stage decisions such as index usage, access paths, and the ordering of operations within a plan. Inaccuracies at this level can be particularly detrimental, as they propagate upward through the query plan, leading to cumulative estimation errors and often resulting in poor join orderings, unnecessary table scans, or misused indexes~\cite{wang2020we, dutt2019selectivity, yang2019deep, lee2023analyzing}.  In contrast, join CE estimates intermediate result sizes across multiple relations and typically builds upon single-table estimates. As a result, inaccuracies in single-table CE can significantly impair join size estimation~\cite{wu2023factorjoin, kim2024asm}. Recent work increasingly integrates refined single-table estimates into join CE models~\cite{zhu2021glue}, highlighting their foundational role. We, thus, focus on the advancement of single-table CE, recognizing its centrality in driving both local and global plan quality.

%Ideally, a cardinality estimation technique should strike an optimal balance among three key factors: accuracy (ability to generate optimal plans), efficiency (with respect to inference latency and training or model-building times), and memory footprint~\cite{cormode2011synopses, zhu2020flat}. Major commercial database systems generally use sampling or synopses for cardinality estimation~\cite{poosala1997histogram}. Synopsis techniques such as histograms approximate the joint frequency distribution of a relation by making assumptions such as uniformity and attribute value independence~\cite{lan2021survey, poosala1996improved}.  Large errors in selectivity estimation result from these assumptions being frequently violated in real-world datasets. It has been observed that open-source and commercial DBMSes routinely produce up to $10^4-10^8$ times estimation errors on queries over a large number of attributes~\cite{yang2019deep, leis2015good}. Even with such large errors, commercial DBMSes still use histograms and sampling methods, as they are fast to construct and cheap to store~\cite{yang2019deep, wu2023factorjoin}.
An ideal CE method should balance three factors: \textbf{accuracy} (in generating optimal plans), \textbf{efficiency} (in terms of inference and training latency), and \textbf{compactness} (in memory usage)~\cite{cormode2011synopses, zhu2020flat}. Commercial database systems predominantly employ synopses or sampling techniques for CE~\cite{poosala1997histogram}. Histogram-based methods, for instance, approximate joint distributions using independence and uniformity assumptions~\cite{poosala1996improved}, which are frequently violated in real-world data. Consequently, estimation errors in the range of $10^4$ to $10^8$ have been observed in both open-source and commercial DBMSes~\cite{leis2015good, yang2019deep}. Despite this, synopses and sampling remain widely used due to their low cost and simplicity~\cite{wu2023factorjoin}.

%With the rise of “ML for DB”, multiple recent papers have shown that learned models can greatly improve the cardinality estimation accuracy compared with synopsis and sampling techniques. Even though learned models are accurate, they often suffer from high training costs, inference costs and hyper-parameter tuning costs. As a result, learned models remain challenging to integrate into real-world DBMSes~\cite{wang2020we}. Therefore, our main objective in this study is to develop a learned cardinality estimator that is closer to ideal. 

%Recent progress in machine learning has led to \textit{learned cardinality estimators}, which significantly outperform traditional methods in accuracy~\cite{han2021cardinality, wang2020we, hasan2020deep}. However, these models often incur high training costs, long inference times, and require substantial hyperparameter tuning~\cite{wang2020we}. These practical limitations hinder their integration into production systems. Thus, the central goal of this work is to design a learned CE approach that better balances accuracy, efficiency, and memory footprint.

Recent advances in machine learning have enabled \textit{learned cardinality estimators}, which significantly outperform traditional methods in accuracy~\cite{han2021cardinality, wang2020we, hasan2020deep}. However, they often incur high training costs, long inference times, and require substantial hyperparameter tuning~\cite{wang2020we}. These practical limitations hinder adoption in production systems. This work aims to design a learned CE approach that more effectively balances accuracy, efficiency, and memory footprint.

%Current state-of-the-art learned methods are comprised of query-driven approaches that focus on learning from the characteristics of the queries themselves~\cite{park2020quicksel, dutt2019selectivity, hasan2020deep, kipf2018learned}, and data-driven approaches that aim to summarize the joint data distribution~\cite{heimel2015self, yang2019deep, hilprecht2019deepdb, wu2021unified}. Query-driven methods offer the advantage of very fast inference times, comparable to traditional methods. However, they require large amounts of training data to accurately capture the joint data distribution. Furthermore, if the distribution of the training queries differs from that of the testing queries, query-driven methods may suffer a significant drop in accuracy, which limits their generalizability~\cite{hilprecht2019deepdb}. Although data-driven methods overcome these limitations, their inference times are significantly higher than those of query-driven methods due to the reliance on Monte Carlo sampling techniques during inference~\cite{yang2019deep, lin2023cardinality}. 
State-of-the-art learned CE methods mainly fall into two paradigms: \textit{query-driven} and \textit{data-driven}. Query-driven models learn from features of training queries~\cite{park2020quicksel, dutt2019selectivity, hasan2020deep, kipf2018learned}, offering fast inference times comparable to traditional methods. However, they typically require large, diverse training query sets to generalize well, and often suffer when the distribution of test queries diverges from that of the training set~\cite{hilprecht2019deepdb}. In contrast, data-driven models learn a representation of the data’s joint distribution~\cite{heimel2015self, yang2019deep, hilprecht2019deepdb, wu2021unified}, yielding higher accuracy and better generalization, but at the cost of slower inference, often due to Monte Carlo sampling~\cite{yang2019deep, lin2023cardinality}.
%Alongside these challenges, a common issue for both approaches is the increase in model size and training time as dataset size grows. However, this growth is significantly more pronounced in data-driven methods. These models aim to capture and condense data information into models to reflect joint distributions. As data volume increases, the complexity of the model will also increase, and training will become more challenging. In contrast, even though the dataset is larger, the model complexity and training of the query-driven models do not necessarily become higher given a fixed number of training queries~\cite{wang2020we, kim2022learned}. It is now clear that query-driven and data-driven methods each have advantages and disadvantages.
A major bottleneck shared by both paradigms is \emph{scalability}. As the data set grows, the size and training time of the model increase - more so for data-driven models, which must learn increasingly complex joint distributions. In contrast, query-driven models maintain manageable complexity given a fixed query workload~\cite{kim2022learned}. This trade-off motivates a hybrid approach.

%Taking every aspect into consideration, we propose CoLSE, a novel learned approach for single table cardinality estimation by utilizing both data and query workload, enabling the model to benefit from the strengths of both data-driven and query-driven paradigms. Unlike previous data-driven models that focus on modeling joint probability density functions (PDF), our approach directly models the joint probability over queried intervals or ranges, thus avoiding the use of expensive Monte Carlo sampling techniques during inference. To achieve this, we developed a new algorithm using copula theories. The key advantage of copulas lies in their ability to construct the joint cumulative distribution function (CDF) based on the marginal CDFs of individual attributes.\newline

%\vspace{0.2em}
\noindent
\textbf{Our Approach.} We propose \textbf{CoLSE}, a novel hybrid learned method for single-table cardinality estimation that leverages both data and query workload. CoLSE combines the accuracy of data-driven models with the inference efficiency of query-driven approaches. While a few hybrid approaches exist, most~\cite{hasan2020deep, alece} treat data and queries separately, lacking unified modeling of the joint data distribution. UAE~\cite{wu2021unified} addresses this limitation but still models joint probability density functions (PDFs) and relies on progressive Monte Carlo sampling, similar to autoregressive models like Naru~\cite{yang2019deep}. In contrast, we reformulate the problem as direct joint CDF computation over query ranges, thereby avoiding expensive Monte Carlo sampling during inference.
To achieve this, we introduce a new copula-based algorithm. Copulas~\cite{Wikipedia} are well-suited for constructing a joint cumulative distribution function (CDF) from the marginal CDFs of individual attributes, offering both accuracy and interpretability.

In summary, we make the following contributions.
\begin{enumerate}[nosep, topsep=1pt, leftmargin=*]
      \item We propose a hybrid approach for single-table cardinality estimation that combines a novel D-vine copula-based algorithm as the data-driven component with a lightweight neural network as the query-driven component. The neural network learns and corrects estimation errors introduced by the data-driven model.
    
       \item We introduce a novel and interpretable algorithm for modeling joint data distributions via marginal CDFs, grounded in D-vine copula theory. To the best of our knowledge, this is the first architecture to incorporate CDF-based joint distribution modeling into single-table cardinality estimation. We further demonstrate that vine copulas can be effectively applied to this task.
    
    %\item We present a new evaluation metric by modifying the PostgreSQL source code to better assess the impact of cardinality estimation methods. While traditional metrics such as Q-error focus solely on the accuracy of individual sub-plan estimates, they fail to capture the end-to-end influence of estimation errors on query performance. To address this limitation, and inspired by recent studies~\cite{9094107, negi2021flow, han2021cardinality}, we propose a combined metric that incorporates both the number of optimal query plans and end-to-end execution time. This composite evaluation offers a more holistic view of the practical impact of cardinality estimation methods on query performance.
        \item We present a new evaluation metric by modifying the PostgreSQL source code to better assess the impact of cardinality estimation methods. Traditional metrics such as Q-error focus solely on local estimation accuracy and overlook broader execution outcomes. Inspired by recent studies~\cite{9094107, negi2021flow, han2021cardinality}, our metric is based on the number of optimal query plans,which has been shown to strongly correlate with runtime, thereby providing a more holistic measure of practical performance.
        
    \item We conduct extensive experiments on both real-world and synthetic datasets. Results show that our model achieves a favorable trade-off among accuracy, training time, inference latency, and model size, outperforming state-of-the-art methods across multiple benchmarks.
\end{enumerate}

%\begin{figure*}
 %   \centering
 %   \includegraphics[width=\linewidth]{figures/research-paper-main-diagrams.png}
 %   \caption{Overall Architecture}
 %   \label{fig:architect}
%\end{figure*}
\vspace{-1em}
\section{Related Work}
\vspace{-0.5em}
Cardinality estimation has led to a wide range of approaches, from traditional statistical techniques to modern learned models. Here, we review classical methods, along with recent query- and data-driven learning-based estimators.

\noindent
\textbf{Traditional methods.} Multidimensional histograms~\cite{poosala1997selectivity, muralikrishna1988equi, aboulnaga1999self, bruno2001stholes} are among the most well-studied techniques for capturing attribute correlations~\cite{dutt2019selectivity}. However, they often require significant storage space to maintain accuracy. Sampling-based approaches~\cite{wu2001applying} can better capture complex correlations and dependencies among attributes. Nonetheless, samples may become stale as the underlying data changes, and sampling methods can incur high storage and retrieval overhead, especially on large datasets~\cite{lan2021survey}.

\noindent
\textbf{Query-driven learned CE methods.} These methods treat CE as a supervised regression task, training models to map queries to estimated result sizes using features extracted from query structures. MSCN~\cite{kipf2018learned} uses a multi-set convolutional network that represents each query as a feature vector composed of table, join, and predicate modules—each implemented as a two-layer neural network. It also leverages a materialized sample to improve learning. LW-XGB and LW-NN~\cite{dutt2019selectivity} propose lightweight models based on XGBoost and neural networks, respectively. Their input features include both range features and CE features, derived from heuristics such as histograms and domain knowledge. DQM-Q~\cite{hasan2020deep} introduces a custom featurization method for training a neural CE model.

\noindent
\textbf{Data-driven learned CE methods.} Data-driven approaches model CE as a joint probability distribution estimation task, aiming to learn the full joint distribution $P(A_1, A_2, \dots, A_n)$ over table attributes, from which selectivity can be inferred. These models are typically unsupervised and rely on deep autoregressive models or probabilistic graphical models (PGMs). Naru~\cite{yang2019deep} and DQM-D~\cite{hasan2020deep} utilize deep autoregressive architectures such as MADE~\cite{germain2015made} and Transformers~\cite{han2021transformer} to approximate conditional probabilities between attributes. However, these models suffer from high training and inference times, limiting their applicability in real-world DBMSs. DeepDB~\cite{hilprecht2019deepdb} employs relational sum-product networks (RSPNs), a type of PGM, to capture both marginal and joint distributions. A key limitation of SPNs is that they retain local independence assumptions~\cite{wang2020we}. Other PGM-based methods use Bayesian networks~\cite{getoor2001selectivity, tzoumas2011lightweight} to model conditional independencies, but structure learning is NP-hard~\cite{zhu2020flat, scanagatta2019survey}, making them expensive to train on large datasets.

Complimentary to the above, there is a line of work which improves query performance by tolerating CE errors via learned steering—learning hint sets or ranking/forcing plans from runtime feedback~\cite{marcus2021bao, woltmann2023fastgres, xu2023coool, zhu2023lero}. In contrast, direct CE estimation improves the estimates themselves, yielding system-wide, interpretable gains without exploration or cold-start overhead and ensuring predictable behavior within standard optimizer abstractions.

\vspace{-0.5em}
\section{Problem Statement}
\label{sec:probstat}

We formally define the task of cardinality estimation for single-table queries involving range and equality predicates.

Consider a relation $T$ consisting of $n$ columns (or attributes), denoted as $\{A_1, A_2, \dots, A_n\}$. A tuple $x \in T$ is an $n$-dimensional vector. A query $q$ is defined as a conjunction of predicates, where each predicate imposes a constraint on a single attribute. Predicates may take the form of an equality constraint ($A_k = c$), an open range constraint ($lb_k \leq A_k$), or a closed range constraint ($lb_k \leq A_k \leq ub_k$).

\noindent \textbf{Cardinality.} The \textit{cardinality} of query $q$, denoted by $|q(T)|$, is the number of tuples in $T$ that satisfy all conditions specified in $q$:
\vspace{-1em}
\begin{equation}
    |q(T)| = \sum_{x \in T} \prod_{k=1}^n I_k(x)
\end{equation}

\noindent
where $I_k(x)$ is an indicator function that evaluates to 1 if tuple $x$ satisfies the predicate on attribute $A_k$, and 0 otherwise.

\noindent \textbf{Selectivity.} The \textit{selectivity} of $q$, denoted as $sel(q)$, is the probability that a randomly selected tuple from $T$ satisfies all predicates in $q$:
\vspace{-0.5em}
\begin{equation}
    sel(q) = \frac{|q(T)|}{|T|}
\end{equation}

\noindent
Alternatively, in probabilistic terms, if each predicate defines a bounded range, the selectivity can be expressed as:
\vspace{-0.5em}
\begin{equation}
    sel(q) = P(lb_1 \leq A_1 \leq ub_1, \dots, lb_n \leq A_n \leq ub_n)
\end{equation}

\noindent \textbf{Objective.} In this work, we focus on modeling \textit{selectivity} rather than cardinality directly. Since the two are linearly related via the total number of tuples $|T|$, modeling selectivity provides a probabilistic foundation that more naturally captures predicate interactions and generalizes across queries. Our goal is to build a selectivity estimation technique that achieves an optimal balance between accuracy, memory efficiency, and inference/training time by leveraging both the underlying data and the observed query workload.
\vspace{-0.5em}
\section{Background on Copula Models}
%\vspace{-0.5em}
% Selectivity estimation fundamentally requires modeling the joint distribution of query attributes. Many traditional approaches assume independence between attributes, which can lead to large estimation errors in the presence of statistical dependencies~\cite{lan2021survey, poosala1997selectivity}. To address this, we leverage \textbf{copula theory}\revision{\info{R4:O1}—a principled framework that constructs the joint distribution by first modeling the marginal distributions of individual attributes and then separately capturing their dependency structure. This decoupling is particularly advantageous for cardinality estimation, as it allows for flexible and accurate representation of complex inter-attribute relationships, including non-linear and asymmetric dependencies, which are common in real-world datasets. We refer readers to~\cite{HudsonThamesVineCopulaIntro, Wicklin2021CopulasIntro, CzadoNagler2022VineCopula} for tutorials on copulas.}

Selectivity estimation fundamentally requires modeling the joint distribution of query attributes. Many traditional approaches assume independence between attributes, which can lead to large estimation errors in the presence of statistical dependencies~\cite{lan2021survey, poosala1997selectivity}. To address this, we leverage \textbf{copula theory}—a mathematical function that captures the dependence between random variables. Its applicability to modeling joint distributions is grounded in \textbf{Sklar’s theorem}~\cite{ling2020deep}, which states that any multivariate joint distribution can be constructed by first modeling the marginal distributions of individual attributes and then separately capturing their dependency structure. This decoupling is particularly advantageous for cardinality estimation, as it allows for flexible and accurate representation of complex inter-attribute relationships, including non-linear and asymmetric dependencies, which are common in real-world datasets. We refer readers to~\cite{HudsonThamesVineCopulaIntro, Wicklin2021CopulasIntro, CzadoNagler2022VineCopula} for more detailed tutorials on copulas.

% \info{This paragraph can merge with the above one as they both introduce copula}
% A copula is a mathematical function that captures the dependence between random variables. Its applicability to modeling joint distributions is grounded in \textbf{Sklar’s theorem}~\cite{ling2020deep}, which states that any multivariate joint distribution can be expressed in terms of its univariate marginal distribution functions and a copula function that encodes their dependencies. 
Formally, for a random vector $[X_1, ..., X_d]$ with marginal CDFs $F_i(x_i)$, the joint CDF can be written as:
\begin{equation}
    P(X_1 \leq x_1, ..., X_d \leq x_d) = C(F_1(x_1), ..., F_d(x_d)),
\end{equation}
where $C$ is the copula function. 
This transformation is enabled by the \textbf{Probability Integral Transform}~\cite{wikipedia_probability_integral_transform}, which states that applying the CDF of a continuous random variable $X$ to itself yields a uniformly distributed variable $U = F_X(X)$ over $[0,1]$. This allows copulas to operate in a normalized domain, making them broadly applicable to multivariate modeling.

Several families of copulas exist, including Gaussian, t-copulas, and Archimedean copulas~\cite{Wikipedia}. While effective in low dimensions, these classical models often struggle to scale due to increasing complexity and rigidity.

\begin{figure}
  \centering
  \includegraphics[width=0.59\linewidth]{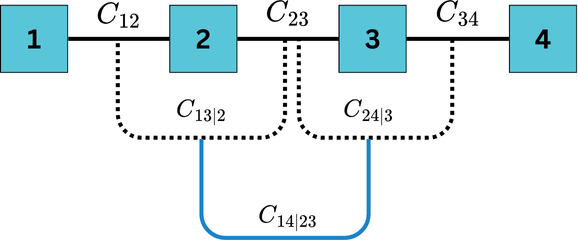}
  \caption{D-vine structure for four variables: dependencies are modeled sequentially.}
  \label{fig:dvine}
   \vspace{-0.5em}
\end{figure}

To overcome these limitations, \textbf{vine copulas} were introduced~\cite{aas2009pair}, enabling the construction of high-dimensional copulas using hierarchical compositions of bivariate (pairwise) copulas. This modular design enhances flexibility and interpretability. Among vine copulas, we adopt the \textbf{D-vine copula} due to its natural alignment with the structure of conjunctive predicates in SQL queries. In a D-vine, variables are arranged in a sequence, and dependencies are modeled step-by-step between adjacent variables (Fig.~\ref{fig:dvine}). This sequential structure simplifies training, mitigates overfitting, and allows the model to scale to higher dimensions while maintaining tractability~\cite{brechmann2013modeling, czado2022vine}.

Therefore, we incorporate D-vine copulas into our model, using them to accurately and efficiently capture the joint cumulative distribution of query attributes. However, classical vine copulas are primarily defined for conditional densities. This requires designing a novel algorithmic adaptation that transforms the traditional density-based formulation into a CDF-based computation suitable for multi-attribute range queries. The details of these algorithmic adaptations and their integration with the query-driven components are presented in the following sections.

\begin{figure*}
   \vspace{-1em}
    \centering
    \includegraphics[width=0.89\linewidth]{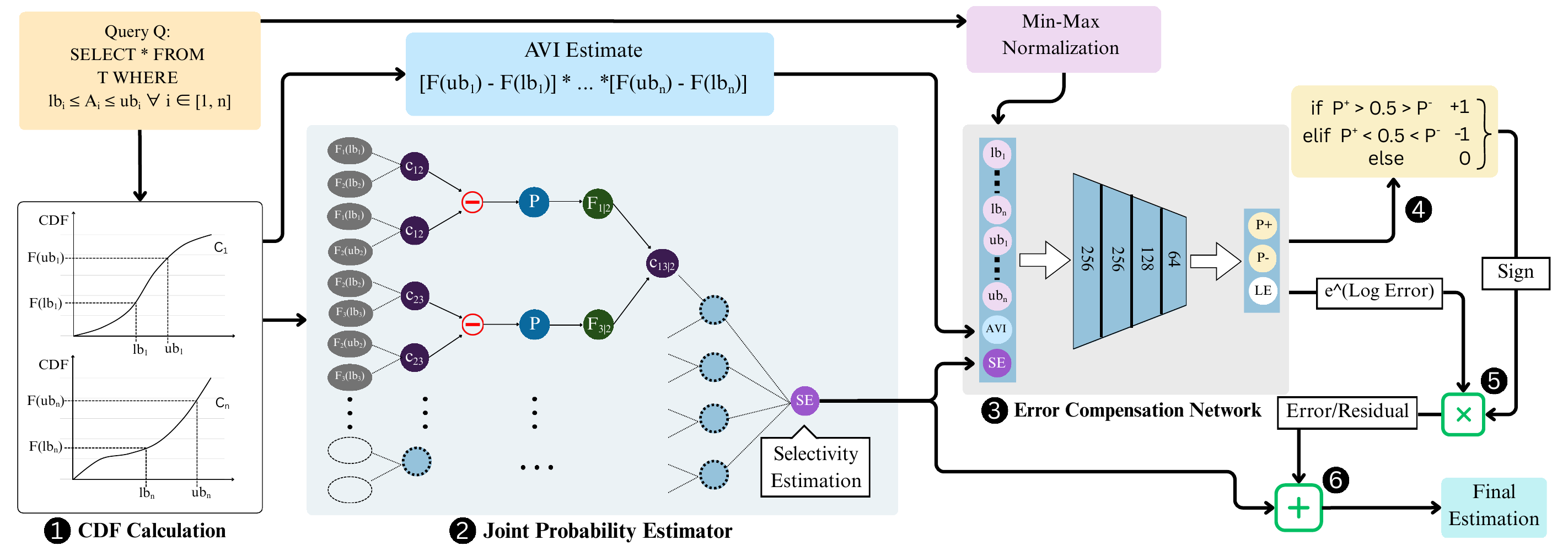}
       \vspace{-0.5em}
    \caption{ CoLSE: The Overall Architecture}
    \label{fig:architect}
     \vspace{-0.5em}
\end{figure*}

\vspace{-0.5em}
\section{C\MakeLowercase{o}LSE Framework}
% \rewrite{
% This section presents the design and core components of the CoLSE framework, a hybrid selectivity estimator that integrates data-driven and query-driven learning. CoLSE models complex attribute dependencies using copula-based decomposition and refines its estimates through a learned error correction mechanism. We begin by providing an overview of the system architecture, then detail the construction of marginal distributions, the D-vine-based joint probability estimator, the error compensation network, and the handling of categorical and discrete attributes.}\\
This section presents the design and core components of the CoLSE framework, a hybrid selectivity estimator that integrates data-driven and query-driven learning. CoLSE models complex dependencies via a copula-based decomposition and refines estimates with a learned error-correction module. We first outline the architecture, then describe marginal distribution construction, the D-vine-based joint probability estimator, the error-compensation network, and support for categorical/discrete attributes and joins.

% \vspace{-0.5em}
% \begin{figure*}
%     \centering
%     \includegraphics[width=\linewidth]{figures/research-paper-main-diagrams.png}
%     \caption{Overall Architecture}
%     \label{fig:architect}
% \end{figure*}
% \begin{figure*}
%     \centering
%     \includegraphics[width=\linewidth]{figures/architecture_1-3.pdf}
%     \caption{Overall Architecture}
%     \label{fig:architect}
% \end{figure*}

\vspace{-0.5em}
\subsection{An Overview}
CoLSE (\underline{Co}pula based \underline{L}earned \underline{S}electivity \underline{E}stimator) is a hybrid cardinality estimator that models the joint data distribution using copula-based decomposition. It integrates two key components:
(1) a joint probability estimator (JPE) that captures the joint data distribution using copula-based decomposition (data-driven), and
(2) an error compensation network (ECN) that uses query workloads to refine the estimates for higher accuracy (query-driven).
CoLSE is categorized as a hybrid approach due to its ability to learn from both data and observed query workloads.

Figure~\ref{fig:architect} illustrates the CoLSE architecture. Given a query $Q$, we begin by extracting the lower and upper bounds ($lb_i$, $ub_i$) for each selection predicate $x_i$, then evaluate their marginal CDFs $F(lb_i)$ and $F(ub_i)$ (Step~\ding{202}). These values are passed to the JPE, a tree-structured module based on D-vine copulas (Step~\ding{203}). Unlike conventional data-driven methods that approximate the joint probability density function (PDF), the JPE directly estimates the joint probability over the query ranges. This avoids costly sampling procedures and leads to significantly lower inference latency.

Next, the output of the JPE, along with the input CDFs, is passed to the ECN, a lightweight neural network trained on past query workloads (Step \ding{204}). The ECN provides targeted corrections to the initial estimate, compensating for inaccuracies arising from modeling limitations or distributional shifts. 

Residual correction is computed based on the ECN output (Steps~\ding{205}–\ding{206}) and applied to the initial estimate, yielding the final selectivity prediction (Step~\ding{207}). This architecture combines the strengths of data- and query-driven methods, resulting in robust and accurate selectivity estimation.

\vspace{-0.5em}
\subsection{Marginal CDF Distribution Modelling}  
%Marginal CDFs are essential inputs for the JPE, forming the basis of copula decomposition. To construct each marginal, we partition the data into $B$ bins (e.g., $B=5{,}000$) and compute the empirical CDF at bin edges. We then fit a Piecewise Cubic Hermite Interpolating Polynomial (PCHIP) to the $(x_j, \text{CDF}_j)$ pairs. This ensures a smooth, strictly monotonic spline bounded in $[0,1]$. At inference time, the marginal CDF of a query point $x$ is efficiently obtained using the PCHIP spline.
Marginal CDFs are essential inputs to the JPE, forming the basis for copula decomposition. To construct each marginal, we partition the data into $B$ bins (e.g., $B = 5{,}000$) and compute the empirical CDF at bin edges. We then fit a Piecewise Cubic Hermite Interpolating Polynomial (PCHIP) to the $(x_j, \text{CDF}_j)$ pairs, yielding a smooth, strictly monotonic spline bounded in $[0,1]$. At inference time, the marginal CDF of a query point $x$ is quickly retrieved using the PCHIP spline.

% \begin{algorithm}[t]
% \begin{small}
% \SetKwInOut{Input}{Input}
% \SetKwInOut{Output}{Output}
% \caption{D-vine Copula Estimation. Note that $\theta$ is needed in all the copula function ($C$) calculations; we omit it due to space limitations.
% }
% \vspace{-0.3em}
% \label{algo:dvine}

% \Input{
%  CDF bounds $[F_{i}(lb_i), F_{i}(ub_i)]$ for queried attributes $[i_1,...,i_n]$; \\
%  % Dependency parameters $\theta$ \;
% }
% \Output{Selectivity Estimate (SE)}
% $n \gets$ number of queried attributes\;
% % \lankadinee{Removed n=0 condition}\\
% \If{$n = 1$}{
%     \Return{$F_i(ub_i) - F_i(lb_i)$}\;
% }{
%     \eIf{$n = 2$}{
%         \Return{$C(F_i(lb_i), F_j(lb_j)) - C(F_i(lb_i), F_j(ub_j)) - C(F_i(ub_i), F_j(lb_j)) + C(F_i(ub_i), F_j(ub_j))$}\;
%     }{
%         $B \gets [(lb,lb), (lb,ub), (ub,ub), (ub,lb)]$ \;
%         $res \gets 0$ \;
%         \For{$k = 0$ \KwTo $3$}{
%             $(b_1, b_2) \gets B[k]$ \;
%             $sign \gets (-1)^k$ \;
%             $C_{i_1(b_1),i_n(b_2)|i_2,...,i_{n-1}} \gets Recursive(b_1, [i_1,...,i_n], b_2)$ \;
%             $res \gets res + sign \cdot C_{i_1(b_1),i_n(b_2)|i_2,...,i_{n-1}}$ \;
%         }
%         \Return{$res$}\;
%     }
% }
% \end{small}
% \end{algorithm}

\begin{algorithm}[t]
\begin{small}
\SetKwInOut{Input}{Input}
\SetKwInOut{Output}{Output}
\SetKwInOut{Uses}{Uses}
\caption{D-vine Copula Estimation. Note that $\theta$ is needed in all the copula function ($C$) calculations; we omit it due to space limitations.
}
\vspace{-0.3em}
\label{algo:dvine}

\Input{
Upper and lower bounds $[(lb_i,ub_i)]$ for queried attributes $[i_1,...,i_n]$; \\
 % Dependency parameters $\theta$ \;
}
\Output{Selectivity Estimate (SE)}
\Uses{
{$\{F_{i_1}, F_{i_2}, … , F_n\} $} \tcp*[r]{\scriptsize CDF functions}
}
$n \gets$ number of queried attributes\;
% \lankadinee{Removed n=0 condition}\\
\If{$n = 1$}{
    \Return{$F_i(ub_i) - F_i(lb_i)$}\;
}{
    \eIf{$n = 2$}{
        \Return{$\begin{aligned}[t]
      & C(F_i(lb_i), F_j(lb_j)) - C(F_i(lb_i), F_j(ub_j)) \\
      & {}- C(F_i(ub_i), F_j(lb_j)) + C(F_i(ub_i), F_j(ub_j));
  \end{aligned}$}
    }{
        $B \gets [(lb_1,lb_n), (lb_1,ub_n), (ub_1,ub_n), (ub_1,lb_n)]$\;
        $res \gets 0$\;
        \For{$k = 0$ \KwTo $3$}{
            $(b^x, b^y) \gets B[k]$\;
            $res \gets res + (-1)^k \cdot Recursive(b^x, [i_1,...,i_{n}], b^y)$\;
        }
        \Return{$res$}\;
    }
}
\end{small}
\end{algorithm}

{%
\linespread{1.15}\selectfont 
\begin{algorithm}[t]
\begin{small}
% \begin{scriptsize}
\SetKwInOut{Input}{Input}
\SetKwInOut{Output}{Output}
\SetKwInOut{Uses}{Uses}
\SetKwFunction{Recursive}{Recursive}
\caption{Recursive Conditional Copula Estimation 
% {\color{red}What is the difference between them.}
}
\label{algo:recursive}

\Input{
{$b^x$, $[i_1,\dots,i_n]$, $b^y$}
}
\Output{
{Conditional Copula $C_{i_1,i_n|i_2...i_{n-1}}$}
}
\Uses{
{$\{F_{i_1}, F_{i_2}, … , F_n\}$}
}
\Recursive{$b^x$, $[i_1,\dots,i_n]$, $b^y$}:

\BlankLine
\If{$n = 3$}{

    $U_1 \leftarrow C_{i_1,i_2}\!\big(F_{i_1}(b^{x}_{i_1}),\,F_{i_2}(ub_{i_2})\big)$\;

    $L_1 \leftarrow C_{i_1,i_2}\!\big(F_{i_1}(b^{x}_{i_1}),\,F_{i_2}(lb_{i_2})\big)$\;

    $P(i_{1(b^{x})}, i_2) \leftarrow U_1 - L_1$\;

    $U_2 \leftarrow C_{i_2,i_3}\!\big(F_{i_3}(b^{y}_{i_3}),\,F_{i_2}(ub_{i_2})\big)$\;

    $L_2 \leftarrow C_{i_2,i_3}\!\big(F_{i_3}(b^{y}_{i_3}),\,F_{i_2}(lb_{i_2})\big)$\;

    $P(i_2, i_{3(b^{y})}) \leftarrow U_2 - L_2$\;

    $P(i_2) \leftarrow F_{i_2}(ub_{i_2}) - F_{i_2}(lb_{i_2})$\;

    $F_{i_1|i_2} \leftarrow P(i_{1(b^{x})},i_2) / P(i_2)$\;
    
    $F_{i_3|i_2} \leftarrow P(i_2, i_{3(b^{y})}) / P(i_2)$\;

    % $P(i_{1(b_1)},i_2) = C_{i_1,i_2}(F_{i_1}(b_{1_{i_1}}), F_{i_2}(ub_{i_2})) - C_{i_1,i_2}(F_{i_1}(b_{1_{i_1}}), F_{i_2}(lb_{i_2}))$\;
    
    % $P(i_2, i_{3(b_2)})$ is computed analogously.\;

    % $P(i_2, i_{3(b_2)}) = C_{i_2,i_3}(F_{i_3}(b_{2_{i_3}}), F_{i_2}(ub_{i_2})) - C_{i_2,i_3}(F_{i_3}(b_{2_{i_3}}), F_{i_2}(lb_{i_2}))$\;

    % $P(i_2) = F_{i_2}(ub_{i_2}) - F_{i_2}(lb_{i_2}) $,
    % $ F_{i_1|i_2} = P(i_{1(b_1)},i_2) / P(i_2)$,
    % $ F_{i_3|i_2} = P(i_2, i_{3(b_2)}) / P(i_2)$\;

    return $C_{i_1,i_3|i_2}(F_{i_1|i_2}, F_{i_3|i_2})$\;
}
\Else{
    $C^{lb}_{i_1,i_{n-1}} \leftarrow 
    \textbf{\textit{Recursive}}(b^{x}, [i_1,\dots,i_{n-1}], lb)$\;

    $C^{ub}_{i_1,i_{n-1}} \leftarrow 
    \textbf{\textit{Recursive}}(b^{x}, [i_1,\dots,i_{n-1}], ub)$\;

    $F_{i_1 \mid i_2,\dots,i_{n-1}} 
    \leftarrow \dfrac{(C^{lb}_{i_1,i_{n-1}} - C^{ub}_{i_1,i_{n-1}})\,\cdot\, P(i_2,\dots,i_{n-2})}
                    {P(i_2,\dots,i_{n-1})}$\;
    
    % $C_{i_1(b_1),i_{n-1}(lb)|i_2,...,i_{n-2}}$ = \textbf{\textit{Recursive}}($b_1$, [$i_1, ..., i_{n-1}$], $lb$)
    
    % $C_{i_1(b_1),i_{n-1}(ub)|i_2,...,i_{n-2}}$ = \textbf{\textit{Recursive}}($b_1$, [$i_1, ..., i_{n-1}$], $ub$)\smallskip

    % $F_{i_1|i_2,...,i_{n-1}} = ((C_{i_1(b_1),i_{n-1}(lb)|i_2,...,i_{n-2}} - C_{i_1(b_1),i_{n-1}(ub)|i_2,...,i_{n-2}}) \times P(i_2,...,i_{n-2})) / P(i_2,...,i_{n-1})$\smallskip

% Recursive copula values for lower/upper bounds of i2
    $C^{lb}_{i_2,i_n} \leftarrow 
    \textbf{\textit{Recursive}}(lb,\,[i_2,\dots,i_n],\,b^{y})$\;

    $C^{ub}_{i_2,i_n} \leftarrow 
    \textbf{\textit{Recursive}}(ub,\,[i_2,\dots,i_n],\,b^{y})$\;

% Conditional distribution of in given i2,...,i_{n-1}
    $F_{i_n \mid i_2,\dots,i_{n-1}} 
    \leftarrow \dfrac{\big(C^{lb}_{i_2,i_n} - C^{ub}_{i_2,i_n}\big)\cdot P(i_3,\dots,i_{n-1})}
                    {P(i_2,\dots,i_{n-1})}$\;
    % $C_{i_2(lb),i_n(b_2)|i_3,...,i_{n-1}}$ = \textbf{\textit{Recursive}}($lb$, [$i_2, ..., i_{n}$], $b_2$)

    % $C_{i_2(ub),i_n(b_2)|i_3,...,i_{n-1}}$ = \textbf{\textit{Recursive}}($ub$, [$i_2, ..., i_{n}$], $b_2$)\smallskip

    % $F_{i_n|i_2,...,i_{n-1}} = ((C_{i_2(lb),i_n(b_2)|i_3,...,i_{n-1}} - C_{i_2(ub),i_n(b_2)|i_3,...,i_{n-1}}) \times P(i_3, ..., i_{n-1})) / P(i_2,...,i_{n-1})$\smallskip

    $P(i_2, i_3, ... i_{n}) = P(i_2, i_3, ... i_{n(ub)}) - P(i_2, i_3, ... i_{n(lb)})$

    return $C_{i_1,i_n|i_2...i_{n-1}}(F_{i_1|i_2,...,i_{n-1}},F_{i_n|i_2,...,i_{n-1}})$\;
}
\end{small}
% \end{scriptsize}
\vspace{-0.5em}
\end{algorithm}
}

\vspace{-0.5em}
\subsection{Novel Algorithm based on D-vine Copula}
\label{sec:novelalgo}
At the core of CoLSE’s JPE is a novel algorithm that estimates multi-dimensional selectivities directly over query ranges by combining principles from probability theory and D-vine copulas. 
Unlike prior methods that model the full joint PDF, our approach entirely avoids high-dimensional density estimation by relying solely on \textit{pairwise copula functions}, 
which capture dependencies between attribute pairs by first transforming each attribute to its marginal CDF (i.e., uniform scores), and then modeling their joint behavior in the uniform space. 
This formulation significantly reduces model complexity while preserving essential dependency structures.

In practice, we instantiate each pair-copula using the Gumbel~\cite{genest1993statistical} copula, a member of the Archimedean family known for its simple closed-form expression and single parameter that controls dependence strength. Compared to alternatives such as the Clayton or Frank copulas, the Gumbel copula offers greater flexibility in capturing asymmetric dependencies, which commonly arise in real-world datasets. The dependence parameter $\theta$ is estimated using Kendall’s rank correlation~\cite{gorecki2016approach}, computed from the data points in each attribute pair.

JPE requires a predefined attribute sequence to model dependencies sequentially. The natural schema order is used as a fixed global ordering for all queries. When a query includes only a subset of attributes, their relative positions in the global order are preserved, and intermediate attributes are skipped (e.g., [City, Year, Make, Model] → [Make, Model]). Further justification for this choice is provided in Section~\ref{subsec:sensitivityanalysis}.

Algorithm~\ref{algo:dvine} presents the generalized procedure for selectivity estimation over a table with $n$ query attributes. 
The algorithm is dynamic and adapts its structure based on the number of queried attributes. We distinguish three cases: (i) single-attribute queries, (ii) two-attribute queries, and (iii) multi-attribute queries ($n > 2$).

\textbf{Case 1.}
For a single predicate on attribute $X$, the selectivity reduces to the difference of marginal CDF values (line 3 in Algorithm~\ref{algo:dvine}): $P(lb \leq X \leq ub) = F(ub) - F(lb)$.

\textbf{Case 2.} For two-attribute queries, we apply the inclusion–exclusion principle~\cite{sane2013inclusion} (line 6 in Algorithm~\ref{algo:dvine}). This is because range queries define axis-aligned rectangles in the attribute space, while the joint CDF provides cumulative probabilities from the origin to a point. To compute the probability of a bounded region, we must combine CDF values at the rectangle’s corners while correcting for overlaps. Let $F_1$ and $F_2$ be the marginal CDFs of $X_1$ and $X_2$, respectively. Their joint CDF, $F_{12}$, is approximated via a bivariate copula function $C_{12}$ as: $F_{12}(x_1, x_2) = C_{12}(F_1(x_1), F_2(x_2))$.

The final selectivity estimate is obtained by applying inclusion–exclusion over the rectangular query range (as in Eq.~\eqref{eq:6}).
%\vspace{-0.5em}
{\small
\setlength{\abovedisplayskip}{2pt}
\begin{align} \label{eq:6}
& P(\text{lb}_1 \leq X_1 \leq \text{ub}_1, \text{lb}_2 \leq X_2 \leq \text{ub}_2) \notag \\
& = F_{12}(\text{lb}_1,\text{lb}_2) - F_{12}(\text{lb}_1,\text{ub}_2) - F_{12}(\text{ub}_1,\text{lb}_2) + F_{12}(\text{ub}_1,\text{ub}_2) \notag \\
& = C_{12}(F_1(\text{lb}_1),F_2(\text{lb}_2)) - C_{12}(F_1(\text{lb}_1),F_2(\text{ub}_2)) \notag \\
& \quad - C_{12}(F_1(\text{ub}_1),F_2(\text{lb}_2)) + C_{12}(F_1(\text{ub}_1),F_2(\text{ub}_2)) 
\end{align}
}
%\vspace{-0.5em}

\noindent

In Eq.~\eqref{eq:6}, 
$F_{12}(\text{ub}_1,\text{ub}_2)$ represents the probability that both $X_1$ and $X_2$ are less than or equal to the upper bounds $\text{ub}_1$ and $\text{ub}_2$.  
$F_{12}(\text{lb}_1,\text{ub}_2)$ subtracts the region where $X_1 < \text{lb}_1$, and  
$F_{12}(\text{ub}_1,\text{lb}_2)$ subtracts the region where $X_2 < \text{lb}_2$.  
$F_{12}(\text{lb}_1,\text{lb}_2)$ adds back the region that was subtracted twice by the previous two terms.

\textbf{Case 3.}  
For queries involving more than two attributes,  
we apply the inclusion–exclusion principle to extract the four corner points of the query attributes (line 8 in Algorithm~\ref{algo:dvine}), and derive four signed copula terms corresponding to the outermost attribute pair in the D-vine structure.  
For each copula term, we recursively compute conditional copula values (line 13 in Algorithm~\ref{algo:dvine})  
on the remaining dimensions using Algorithm~\ref{algo:recursive}.  
The recursion progresses by conditioning on intermediate attributes and terminates when it reaches the base case (i.e., $n=3$).

To clarify the recursive computation in Algorithms 1 and 2, we include a worked example for 4 attributes with variable order $A_1, A_2, A_3, A_4$. Consider the example query: 
\texttt{SELECT * FROM T WHERE 10 $\le$ A$_1$ $\le$ 20 AND 5 $\le$ A$_2$ $\le$ 15 AND 100 $\le$ A$_3$ $\le$ 200 AND 50 $\le$ A$_4$ $\le$ 120;} with marginal CDFs $F_1$:[0.2,0.4], $F_2$:[0.3,0.6], $F_3$:[0.1,0.5], $F_4$:[0.25,0.55]. We aim to compute $P(10\leq A_1\leq20, 5\leq A_2\leq15, 100\leq A_3\leq200, 50\leq A_4\leq120) = P(A_1,A_4\mid A_2,A_3)\times P(A_2,A_3)$.\\
\textbf{Step 1 (Level 1):} Compute pairwise probabilities using inclusion-exclusion on $C_{12},C_{23},C_{34}$. Example: $P(A_1,A_2) = C_{12}(0.4,0.6) - C_{12}(0.4,0.3) - C_{12}(0.2,0.6) + C_{12}(0.2,0.3)$.\\
\textbf{Step 2 (Level 2):} Compute conditional CDFs and copulas: $F_{1\mid2},F_{3\mid2},F_{4\mid3}\rightarrow$ then evaluate $C_{13\mid2}$ and $C_{24\mid3}$. Using these, derive $F_{1\mid23}$ and $F_{4\mid23}$.\\
\textbf{Step 3 (Level 3):} Apply inclusion-exclusion on $(A_1, A_4)$ corners using $C_{14\mid23}$ to obtain $P(A_1,A_4\mid A_2,A_3)$, thus the final probability.

\vspace{-0.5em}
\subsection{Error Compensation Network}
\label{sec:errnet}

To further refine the selectivity predicted by the JPE, CoLSE employs a lightweight neural network trained to estimate the residual error. This network corrects systematic biases that are not captured by the copula-based model.

\textbf{Inputs.} The network takes as input: (i) normalized lower and upper bounds of the queried predicates, (ii) the JPE output, and (iii) a heuristic estimate computed via the Attribute Value Independence (AVI) assumption—that is, the product of marginal CDF differences. Following prior work~\cite{dutt2019selectivity}, we find that including AVI improves both robustness and accuracy.

\textbf{Outputs.} The model predicts: (i) the log absolute residual, (ii) the probability that the residual should be added ($P^+$), and (iii) the probability that it should be subtracted ($P^-$). Residual correction is applied only when one probability exceeds the other and is greater than 0.5. ECN architecture separates magnitude and sign predictions to stabilize training by mitigating oscillations. Independent sign heads act as a confidence-driven gate, suppressing uncertain corrections to produce a more reliable and conservative adjustment process.

\textbf{Architecture.} The model consists of four fully connected layers with 256, 256, 128, and 64 neurons, each followed by ReLU activation. The output layer has 3 neurons. This compact architecture enables efficient inference while capturing non-linear patterns between query features and estimation errors.

\textbf{Training.} The model is trained using a custom loss function that combines Mean Squared Error (for residual magnitude) and Binary Cross-Entropy with logits (for residual sign). The target residual is computed as the log difference between the ground truth and the JPE estimate.
 \vspace{-0.5em}
\subsection{Handling Categorical and Discrete Variables}
\label{subsec:catvarhandling}
Copula models and CDFs are inherently defined over continuous domains. To support categorical and discrete variables, we adopt a dequantization technique inspired by~\cite{wang2021face}, which maps discrete values into a continuous space.

For categorical variables, we first order them alphabetically and then apply label encoding (e.g., $E(\text{Cook}) \rightarrow 0$) to convert them into discrete numeric form (see Section~\ref{subsec:sensitivityanalysis} for further discussion). Both categorical and inherently discrete variables are then dequantized using a spline-based continuous distribution. This process involves two steps: (i) constructing a smooth CDF using PCHIP splines, and (ii) sampling from this CDF and inverting the spline via a fast, precomputed lookup table to produce continuous-valued representations.

After transformation, equality predicates on categorical attributes are translated into range queries. For example, if ``Cook'' is mapped to 0, the corresponding predicate becomes $0 \leq x < 1$, covering the continuous interval assigned to that category. This reformulation ensures compatibility with the copula-based estimation process.

\vspace{-0.5em}

\subsection{Extension to Joins}
Extending beyond single tables, we handle equi-joins by treating the join domain as a common key space across all joined tables and estimate the contribution of each key value to the join result.
% Thus far, our approach focused on single-table cardinality estimation. We now extend it to inner equi-joins. The key idea is to treat the join domain as a common key space across all participating relations and estimate the contribution of each key value to the join result. 
Instead of materializing per-key frequencies, which is infeasible for large domains, we partition the key space into bins and approximate within-bin frequencies using averages derived from our copula-based single-table estimator.

For each bin, we compute an approximate density of tuples per relation, and then estimate the join size by combining these densities across relations. 
Summing the contributions of all bins yields the final join cardinality estimate. 
This strategy preserves the advantages of our single-table approach, while scaling naturally to multi-relation equi-joins.

\vspace{-0.5em}
\section{Experimental Setup}
The experiments are conducted on two Linux servers. Model training is performed on a server equipped with 16 Intel Xeon Platinum 8562Y+ CPUs @ 2.80GHz, one NVIDIA L40S-24Q GPU, and 116 GB of memory. End-to-end evaluation on PostgreSQL is carried out on another server with 4 AMD EPYC 7763 CPUs @ 2.50GHz.

\begin{table}[tb]

\caption{Dataset Characteristics}
\vspace{-1em}
\begin{center}
\begin{tabular}{|l|c|c|c|c|}
\hline
\textbf{Dataset} & \textbf{Size (MB)}& \textbf{Rows} & \textbf{Cols/Cat}& \textbf{Domain} \\
\hline
Census & 4.8 & 49K & 13/8 & $10^{16}$\\
\hline
Forest & 44.3 & 581K & 10/0 & $10^{27}$\\
\hline
Power & 110.8 & 2.1M & 7/0 & $10^{17}$\\
\hline
DMV & 972.8 & 11.6M & 11/10 & $10^{15}$\\
\hline
\end{tabular}
\label{tab:datachar}
\end{center}
 \vspace{-1em}
\end{table}
 \vspace{-0.5em}
\subsection{Datasets}
%\vspace{-0.5em}
We evaluate all baselines on \textbf{four} real-world datasets, following the benchmark in~\cite{wang2020we}, and \textbf{two} synthetic datasets. 
% \revision{\info{R1O1} We add xxx datasets with multiple tables...}
Table~\ref{tab:datachar} summarizes the real-world dataset characteristics: “Cols/Cat” denotes the total and categorical column counts, while “Domain” refers to the product of the distinct values in each column. 
% The synthetic datasets are: (1) one with uniform pairwise correlation coefficients between 0.2 and 0.8; and (2) one derived from the TPC-H \texttt{lineitem} table~\cite{fzirak2025tpchskew}, varying the Zipfian skew from 1 to 4 and dataset size from 0.1GB to 20GB (0.1, 1, 10, 20) {\color{red}, which follows \cite{}}.

The synthetic datasets are designed to test robustness and scalability. The first consists of data with uniform pairwise correlation coefficients between 0.2 and 0.8, aimed at evaluating how query-driven methods tolerate changes in data correlation. The second is derived from the TPC-H \texttt{lineitem} table, where we vary the Zipfian skew from 1 to 4 and dataset sizes from 0.1GB to 20GB (0.1, 1, 10, 20), following~\cite{fzirak2025tpchskew}. This setup assesses how data-driven methods scale in training time as dataset size increases.
To test robustness further, we modify the DMV dataset by incrementally inserting 20\% of random, correlated, and skewed data.

\vspace{-0.5em}
\subsection{Query Workloads} 
%\vspace{-0.5em}
The queries are generated using the workload generator from~\cite{wang2020we}. Each query is created in three steps: selecting predicate attributes, choosing query centers, and assigning operators and widths. The query center distribution (e.g., uniform or skewed) is varied to evaluate baseline robustness under different workloads.

Each dataset includes 100,000 training, 10,000 validation, and 10,000 test queries with ground truth cardinalities. For training query-driven baselines, we use 100,000 training and 10,000 validation queries. In contrast, the proposed model’s error compensation network is trained with only 80,000 queries. All models are evaluated on the same 10,000-query test set.

\vspace{-0.5em}
\subsection{Baseline Methods} We compare with eight baselines including three traditional methods and five learned methods.

% \begin{figure*}[htbp]
%     \centering
%     \includegraphics[width=0.95\linewidth]{figures/sota_models.png}
%     \caption{Comparison against State-of-the-Art across four real-world datasets: Census, Forest, Power and DMV, with respect to (a) No. of optimal query plans(\%) (b) Inference time(ms) (c)Training time (min) and (d) Model size(MB)}
%     \label{fig:sota}
% \end{figure*}

\begin{figure*}[htbp]
     \centering
    \begin{subfigure}[b]{0.4\linewidth}
        \centering
        \includegraphics[width=\linewidth]{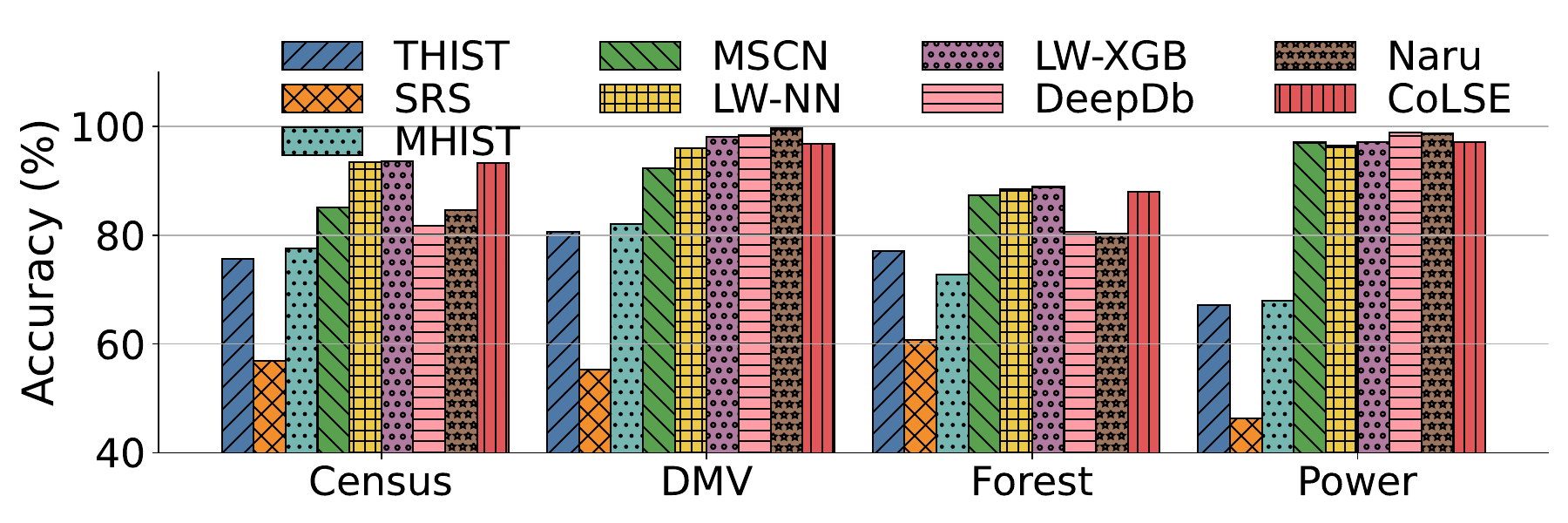}
        \vspace{-1.8em}
        \caption{Accuracy}
        \label{fig:SOTA_accuracy_data}
    \end{subfigure}
    % \hfill
    \begin{subfigure}[b]{0.4\linewidth}
        \centering
        \includegraphics[width=\linewidth]{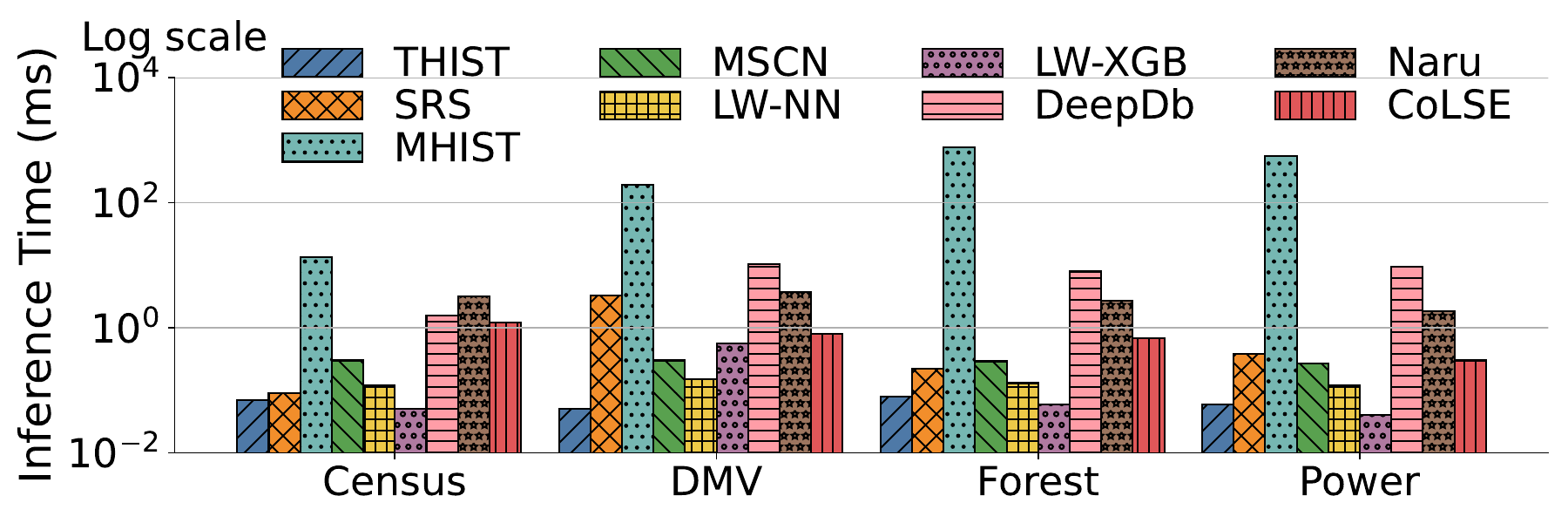}
        \vspace{-1.8em}
        \caption{Inference time (ms)}
        \label{fig:SOTA_inference_time_data}
    \end{subfigure}
    \\
    \begin{subfigure}[b]{0.4\linewidth}
        \centering
        \includegraphics[width=\linewidth]{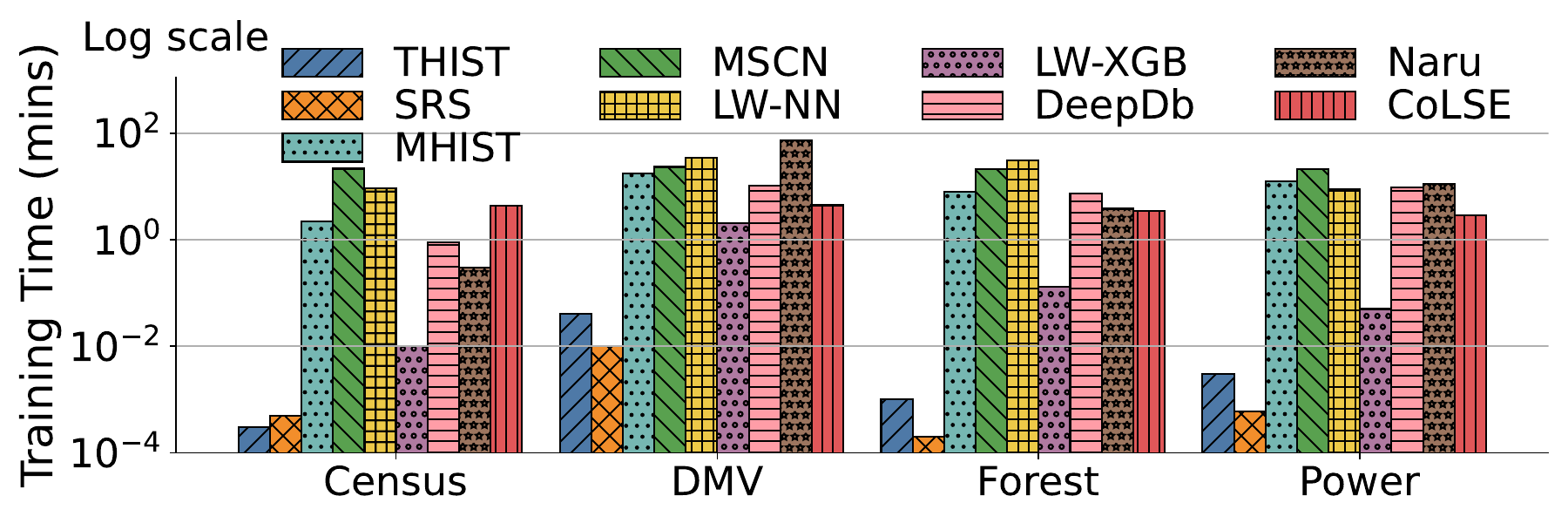}
        \vspace{-1.8em}
        \caption{Training time (min)}
        \label{fig:SOTA_training_time}
    \end{subfigure}
    % \hfill
    \begin{subfigure}[b]{0.4\linewidth}
        \centering
        \includegraphics[width=\linewidth]{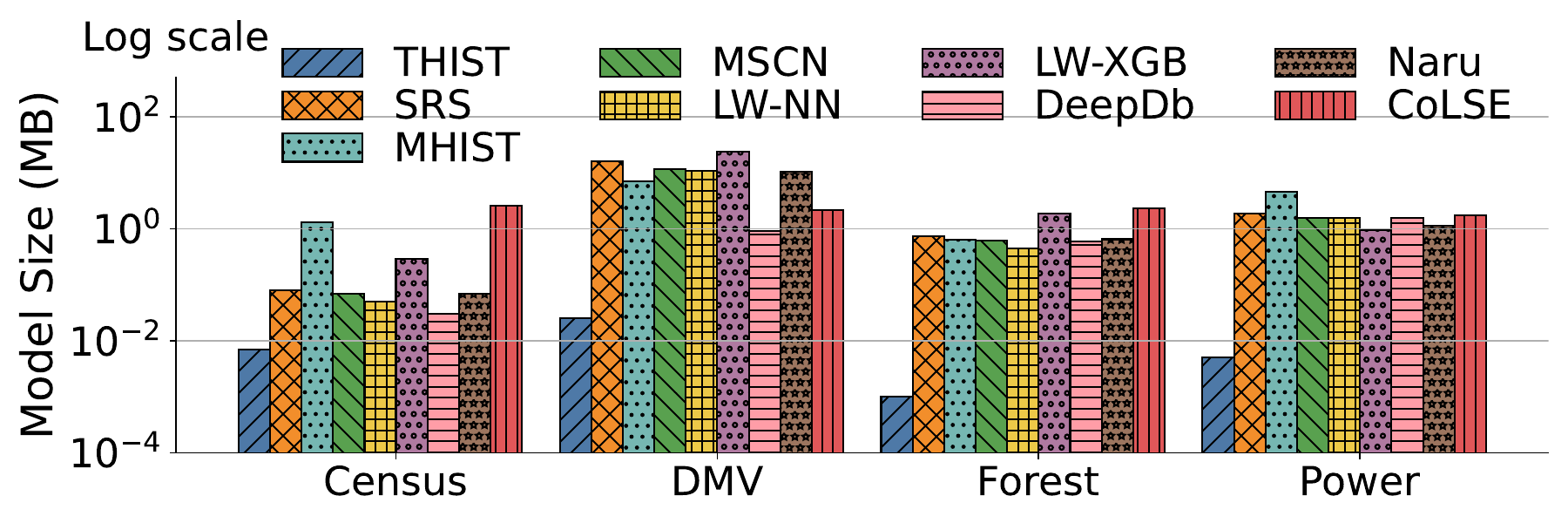}
        \vspace{-1.8em}
        \caption{Model size (MB)}
        \label{fig:SOTA_model_size_data}
    \end{subfigure}
    
    \caption{Comparison against State-of-the-Art across four real-world datasets}
    \label{fig:sota}
    % \vspace{-1.8em}
\end{figure*}

% \textbf{Traditional methods:} 
% \begin{enumerate}[label=\arabic*.]
% \item THIST~\cite{postgresqlrowestimationexamples}: histogram assuming independence.
% \item SRS~\cite{lan2021survey}: simple random sampling.
% \item MHIST~\cite{poosala1997selectivity}: multidimensional histogram.
% \end{enumerate}

% \textbf{Query-driven Models:} 
% \begin{enumerate}[label=\arabic*., resume]
% \item MSCN~\cite{kipf2018learned}: Uses a multi-set convolutional neural network with materialized sampled tuples as additional inputs.
% \item LW-NN~\cite{dutt2019selectivity}: Uses a lightweight neural network.
% \item LW-XGB~\cite{dutt2019selectivity}: Uses a gradient boost tree.
% \end{enumerate}

% \textbf{Data-driven Models:}
% \begin{enumerate}[label=\arabic*., resume]
% \item DeepDb~\cite{hilprecht2019deepdb}: Uses Sum Product Networks (SPN).
% \item Naru~\cite{yang2019deep}: Uses deep autoregressive model, ResMADE.
% \end{enumerate}

\textbf{Traditional methods:} 
\begin{enumerate}[label=\arabic*.]
\item THIST~\cite{postgresqlrowestimationexamples}: a one-dimensional histogram estimator assuming full independence across attributes.
\item SRS~\cite{lan2021survey}: a simple random sampling-based estimator that evaluates selectivity by probing sampled tuples, empirically capturing correlations.
\item MHIST~\cite{poosala1997selectivity}: a multidimensional histogram that partitions the joint data space into buckets to approximate correlated selectivity.
\end{enumerate}

\textbf{Query-driven Models:} 
\begin{enumerate}[label=\arabic*., resume]
\item MSCN~\cite{kipf2018learned}: a supervised neural estimator using multi-set convolutional networks with separate embeddings for tables, joins, and predicates. 
\item LW-NN~\cite{dutt2019selectivity}: a lightweight neural network model trained on handcrafted query features, including predicate ranges and auxiliary CE estimates. 
\item LW-XGB~\cite{dutt2019selectivity}: a gradient-boosted tree model using the same features as LW-NN.
\end{enumerate}

\textbf{Data-driven Models:}
\begin{enumerate}[label=\arabic*., resume]
\item DeepDb~\cite{hilprecht2019deepdb}: a data-driven estimator based on relational sum-product networks (SPNs) that model joint and marginal distributions.
\item Naru~\cite{yang2019deep}: a deep autoregressive model that learns the full joint distribution using Residual Masked Autoencoder for Distribution Estimation (ResMADE) style networks.
\end{enumerate}

%LW-NN and LW-XGB incorporate heuristic features derived from histograms and domain knowledge, improving estimation accuracy.
%We exclude DQM~\cite{hasan2020deep} as its data-driven model performs similarly to Naru, and its query-driven model lacks support for range queries, making it incompatible with our workloads~\cite{wang2020we}.
LW-NN and LW-XGB use heuristic features from histograms and domain knowledge to improve estimation. We exclude DQM~\cite{hasan2020deep} as its data-driven model performs similarly to Naru, while its query-driven model does not support range queries, making it incompatible with our workloads~\cite{wang2020we}.

\vspace{-0.5em}
\subsection{Evaluation Metrics}

\textbf{Accuracy:}
% We measure accuracy by comparing query plans generated using estimated cardinalities against those generated using true cardinalities. For 10,000 test queries, we first obtain ground-truth plans by executing each query with true cardinalities in PostgreSQL. The same queries are then re-executed using estimated cardinalities to produce alternative plans. Accuracy is reported as the proportion of queries where both plans are identical. 
We measure accuracy as the proportion of queries whose plans, generated using estimated cardinalities, are identical to those generated with true cardinalities in PostgreSQL over 10,000 test queries.
We also observed in our experiments that this metric is strongly correlated with end-to-end execution times, further supporting its validity as a practical performance indicator.

\textbf{Inference Latency:}
Since cardinality estimators are invoked repeatedly during query planning, high inference latency slow optimization and negate the benefits of accurate estimates.

\textbf{Model Size:}
Models are often loaded into memory during planning; smaller ones reduce memory use and support deployment in resource-limited settings.

\textbf{Offline Training Time:}
Shorter training times reduce computational costs, particularly for large-scale databases, and facilitate timely model updates in response to data changes.

\textbf{Discussion:}  
%While Q-error is a popular metric for assessing the accuracy of cardinality estimates, it raises the question of whether estimates must be numerically exact. The query optimizer’s true goal is to produce plans that minimize execution time—not to predict cardinalities with perfect precision. Prior work~\cite{9094107, han2021cardinality} has echoed this concern, advocating for alternative metrics that directly reflect query performance, thus shifting the focus from estimate accuracy to end-to-end effectiveness.
Q-error is a common metric for evaluating cardinality estimates, but it raises the question of whether numerical precision is necessary. Since the optimizer’s goal is to minimize execution time--not to predict exact cardinalities--prior work~\cite{9094107, han2021cardinality} has argued for metrics that better reflect query performance, shifting focus from estimate accuracy to overall effectiveness.

% Motivated by the approaches proposed in these references, we introduce a our own evaluation metric to assess the accuracy of both baseline methods and our model.

% \lankadinee{\subsection{Handling data updates}

% Data updates are handled in suc}

\vspace{-0.5em}
\subsection{Modifications to PostgreSQL}  
To evaluate estimation accuracy, we modified PostgreSQL~\cite{PostgreSQL13Documentation} to integrate externally predicted cardinalities. Specifically, we added a function to \texttt{costsize.c}, which computes the cost of potential access paths~\cite{Suzuki2024CostEstimation}, to load selectivities.  
During query planning, PostgreSQL invokes \texttt{set\_baserel\_size\_estimates()} to estimate base relation cardinalities. We extended this function to populate a new field, \texttt{custom\_selectivity}, in the \texttt{PlannerInfo} structure with external values.  
We also modified \texttt{clauselist\_selectivity()}, used throughout \texttt{costsize.c}, to prioritize \texttt{custom\_selectivity} when available. This ensures consistent use of external estimates across all access path evaluations and avoids conflicts between internal and injected values~\cite{Suzuki2024CostEstimation}.

\vspace{-0.5em}
\section{Experimental Results}

To evaluate the effectiveness of CoLSE, we conduct a comprehensive set of experiments addressing the following research questions:

\noindent\textbf{Q1:} How does CoLSE compare to traditional, data-driven, and query-driven baselines in terms of accuracy, inference time, training time, and model size? (Section~\ref{subsec:sota})
\noindent\textbf{Q2:} How do varying column correlations affect these core metrics across all methods? (Section~\ref{subsec:corr})
\noindent\textbf{Q3:} How do dynamic data updates impact model performance and robustness? (Section~\ref{subsec:data_update})
\noindent\textbf{Q4:} How well do different methods adapt to workload shifts? (Section~\ref{subsec:workload_shift})
\noindent\textbf{Q5:} How does data skew influence accuracy and latency across methods? (Section~\ref{subsec:skew})
\noindent\textbf{Q6:} How sensitive are different methods to changes in dataset cardinality? (Section~\ref{subsec:dataset_size})

\vspace{-0.5em}
\subsection{Comparison against State-of-the-Art}
\label{subsec:sota}
% \vspace{-0.5em}
Fig.~\ref{fig:sota} presents the (a) accuracy, (b) inference time, (c) training time, and (d) model size of CoLSE and baseline methods across four real-world datasets.

\begin{figure*}[h!]
     \centering
    \begin{subfigure}[b]{0.4\linewidth}
        \centering
        \includegraphics[width=\linewidth]{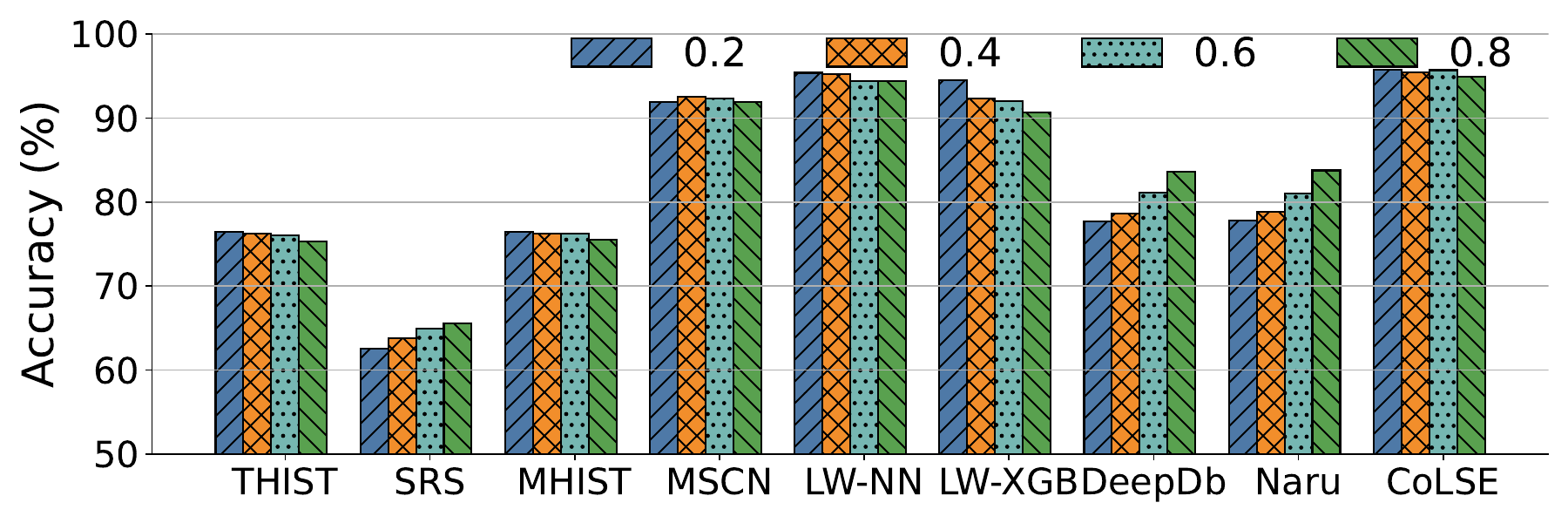}
        \vspace{-1.8em}
        \caption{Accuracy}
        \label{fig:correlation_accuracy}
    \end{subfigure}
    % \hfill
    \begin{subfigure}[b]{0.4\linewidth}
        \centering
        \includegraphics[width=\linewidth]{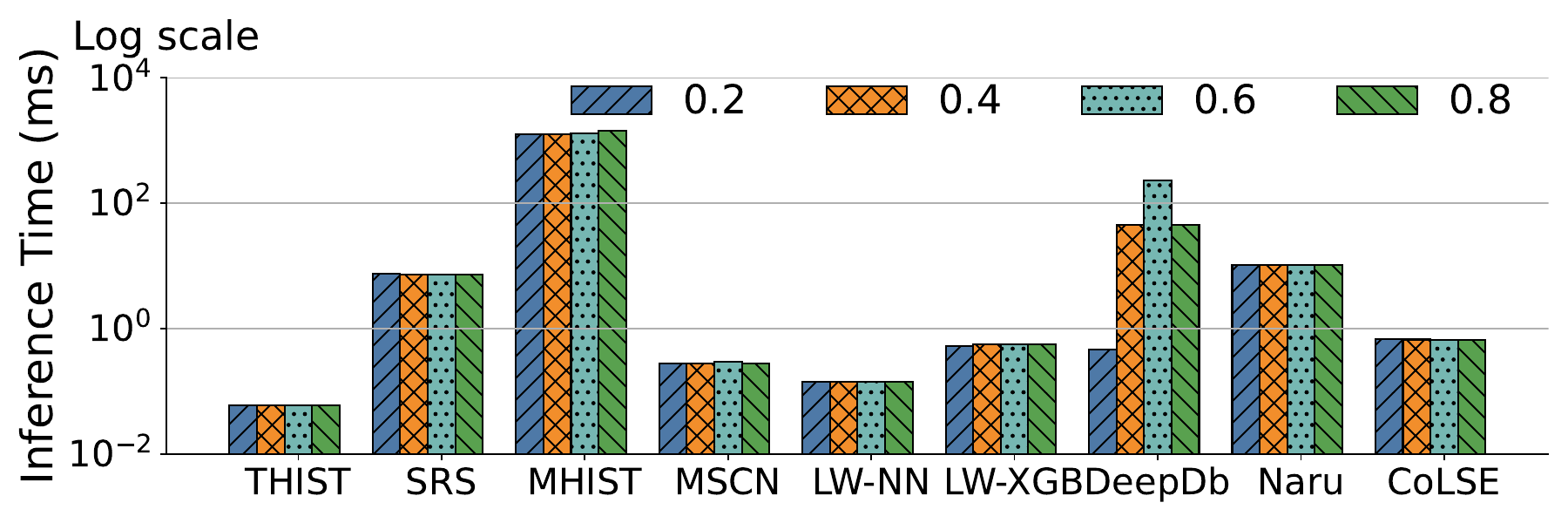}
        \vspace{-1.8em}
        \caption{Inference time (ms)}
        \label{fig:correlation_inference_time}
    \end{subfigure}
    \\
    \begin{subfigure}[b]{0.4\linewidth}
        \centering
        \includegraphics[width=\linewidth]{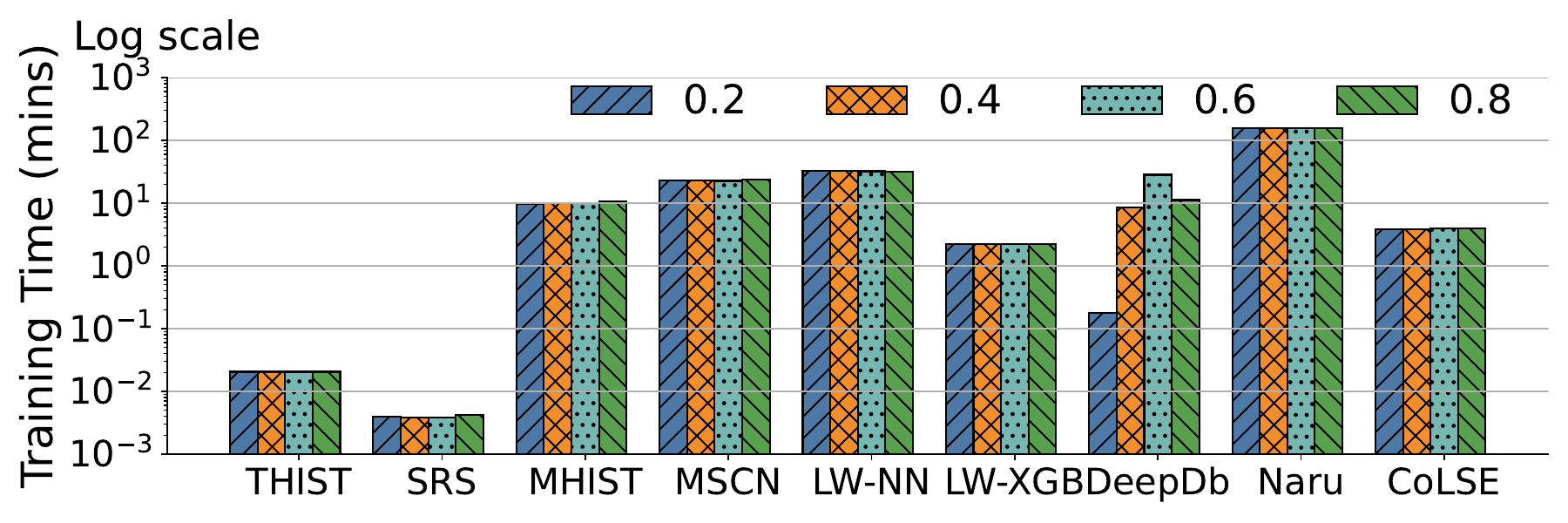}
        \vspace{-1.8em}
        \caption{Training time (min)}  \label{fig:correlation_training_time}
    \end{subfigure}
    % \hfill
    \begin{subfigure}[b]{0.4\linewidth}
        \centering
        \includegraphics[width=\linewidth]{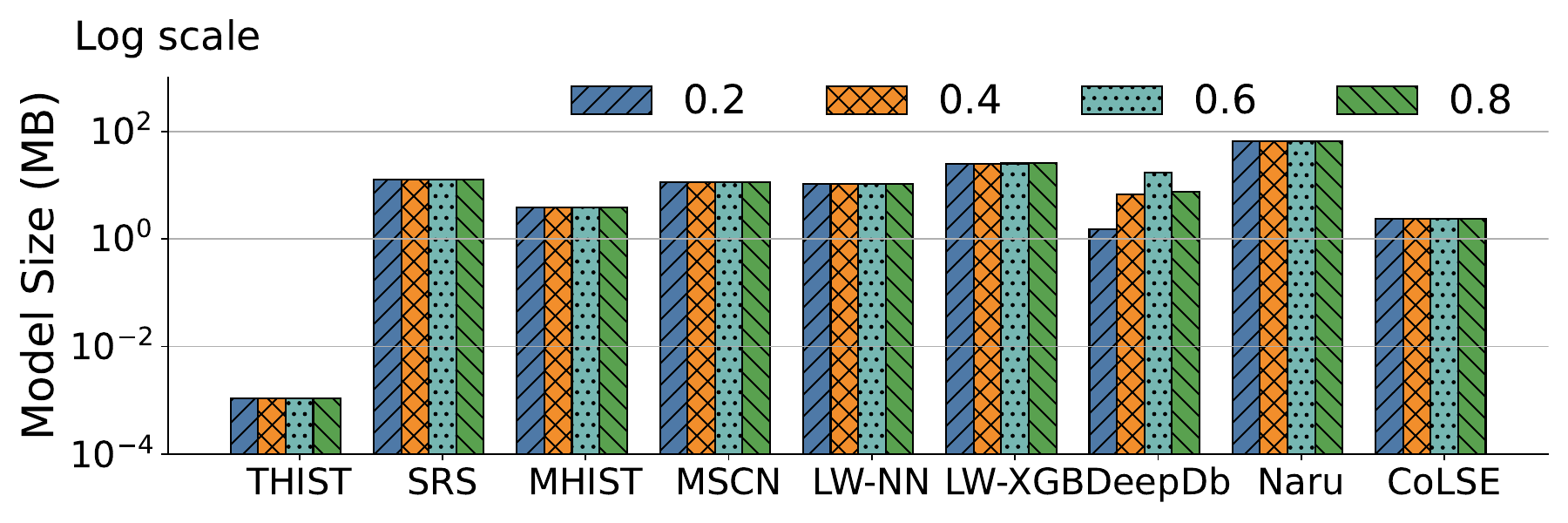}
        \vspace{-1.8em}
        \caption{Model size (MB)}
        \label{fig:correlation_model_size}
    \end{subfigure}
    \vspace{-0.5em}
    \caption{Varying correlation levels}
    \label{fig:corr}
    % \vspace{-1.8em}
\end{figure*}

% \vspace{-0.5em}

\subsubsection{Accuracy Comparison}
CoLSE consistently delivers strong performance across all datasets, matching the accuracy of top-performing models while maintaining robustness.
In contrast, while learned methods generally outperform traditional ones, their performance drops on the Census and Forest datasets. This decline is primarily due to increased randomness in these datasets, which makes it harder for models to learn meaningful patterns. The effect is especially pronounced for data-driven methods, whose accuracy is more dependent on the underlying data structure.

\subsubsection{Inference Latency Comparison}
CoLSE maintains low inference latency (under 1.5 ms), making it suitable for real-time workloads. Although slightly slower than query-driven models due to its two-stage design, it remains significantly faster than data-driven methods.
Among learned models, data-driven approaches show notably higher latency. DeepDB scales poorly, with latency increasing sharply on larger datasets. Naru remains moderate (2–4 ms) but is still slower than query-driven methods like MSCN, LW-NN, and LW-XGB, which all sustain sub-millisecond latencies.
Among traditional methods, THIST offers the lowest latency but limited accuracy; SRS shows moderate and dataset-dependent latency; MHIST, while more accurate, is too slow for real-time use (up to 700 ms).

% \begin{figure*}[htbp]
%     \centering
%     \includegraphics[width=0.95\linewidth]{figures/model_correlation.png}
%     \caption{(a) No. of optimal query plans(\%) (b) Inference time(ms) (c) Training time(min) and (d) Model size(MB)  at varying correlation levels}
%     \label{fig:corr}
%     \vspace{-1em}
% \end{figure*}

\subsubsection{Training Time Comparison}
CoLSE achieves one of the fastest training times among learned models, staying under 5 minutes even on large datasets like DMV. While not as fast as LW-XGB, it outperforms all other query-driven and data-driven baselines in training efficiency, making it practical for deployment.
Among learned baselines, MSCN and LW-NN generally train slower ($>25$ minutes) due to workload encoding and model complexity. LW-XGB is a notable exception, completing training in 2–5 minutes across datasets. 
Data-driven models exhibit high variability: Naru takes over 70 minutes on DMV but only 0.5 minutes on Census. Traditional methods like THIST and SRS are extremely fast (a few seconds), while MHIST, though more accurate, becomes slower on large datasets (15+ minutes on DMV).

\subsubsection{Model Size Comparison}
CoLSE maintains a compact representation—under 3 MB—even on large datasets like DMV, while also achieving low training and inference times. Its model size is mainly determined by the marginal CDFs and parameters of the D-vine and error compensation network.
As shown in Fig.~\ref{fig:SOTA_model_size_data}, model sizes rise significantly for the largest dataset (DMV) across all approaches except DeepDb and CoLSE.
THIST remains the most lightweight among traditional methods. In contrast, MHIST and SRS exhibit model sizes comparable to learned models. Interestingly, DeepDb maintains the smallest model size across all datasets despite higher training and inference costs.
LW-XGB, while efficient in training and inference, incurs a steep model size increase—reaching \raisebox{0.25ex}{\texttildelow}24 MB on DMV—indicating high memory usage and limited portability.

\vspace{-0.3em}
\subsection{Impact of Column Correlation}
\label{subsec:corr}
Fig.~\ref{fig:corr} shows the accuracy, training and inference time, and model size of all under varying column correlations ($\rho$).
\subsubsection{Accuracy Comparison} 
CoLSE is best across all correlations (peak 95.79\%), maintaining top-tier performance. Learned baselines beat traditional methods but generally trail CoLSE in robustness. Among them, query-driven models outperform data-driven ones. DeepDB and Naru show improved accuracy as correlation increases ($\sim$77\% to 83.5\% and 77.8\% to 83.76\%), indicating these models benefit from more predictable attribute patterns. MSCN and LW-NN stay consistently high, whereas LW-XGB degrades at higher correlation (94.5\% at 0.2 to 90.69\% at 0.8), indicating reduced robustness.

% \begin{figure*}[htbp]
%     \centering
%     \includegraphics[width=0.95\linewidth]{figures/model_skew.png}
%     \caption{(a) No. of optimal query plans(\%) (b) Inference time(ms) (c) Training time(min) and (d) Model size(MB) at varying skew levels}
%     \label{fig:skew}
%     \vspace{-1em}
% \end{figure*}

\subsubsection{Inference Latency Comparison}

% Among learned models, CoLSE maintains low inference latency (less than 1 ms) across all correlation levels. Query-driven methods also maintain sub-millisecond latencies, while data-driven models are slower: Naru holds steady around 10 ms, and DeepDb consistently has the highest latency among learned models.

% Among traditional methods, MHIST incurs the highest latency (over 1,200 ms). THIST remains relatively fast, while SRS has moderate latency (\raisebox{0.25ex}{\texttildelow}7 ms), still higher than most learned models.

CoLSE maintains sub-millisecond inference times (\raisebox{0.1ex}{\scriptsize\textless}1 ms)
 across all correlations, comparable to query-driven methods. Data-driven models are slower: Naru holds steady around 10 ms;  while DeepDB is highest among learned approaches.
Among traditional baselines, MHIST is slowest (\raisebox{0.1ex}{\scriptsize\textgreater}1{,}200\,ms). THIST is relatively fast, and SRS is moderate ($\sim$7 ms), still above most learned models.

Most models show stable latency across correlations because hyperparameters were fixed to isolate correlation effects. DeepDB is the exception: despite identical settings, its latency is non-monotonic—rising to ($\sim$220 ms) at $\rho = 0.6$ and then falling to ($\sim$45 ms) at $\rho = 0.8$—likely due to how its RSPN ensemble is constructed from attribute correlations.

% The consistency in inference times across all correlation levels for most models can be attributed to the use of identical model parameters throughout the experiments. This design choice was made to isolate the impact of correlation on model performance. However, DeepDb presents a notable exception. Despite using the same parameters, it exhibits a non-monotonic trend: inference time increases significantly at moderate correlation (peaking around 220 ms at 0.6) and then drops at higher correlation levels (to approximately 45 ms at 0.8). This is likely due to the way DeepDb constructs its ensemble of Relational Sum-Product Networks (RSPNs) based on attribute correlations.

\subsubsection{Training Time Comparison}
% Training time trends generally mirror inference latency patterns. CoLSE is among the fastest learned models, staying below 5 minutes, second only to LW-XGB. Naru is the slowest, reaching 	\raisebox{0.25ex}{\texttildelow}160 minutes. DeepDb again shows a spike at correlation 0.6 (\raisebox{0.25ex}{\texttildelow}30 minutes), consistent with its inference time behavior.
% MSCN and LW-NN fall in the mid-range (20–30 minutes), while LW-XGB trains in under 5 minutes. As expected, traditional methods require minimal training time: MHIST completes in under 10 minutes, and THIST and SRS are built within seconds.
Training time trends generally mirror inference latency patterns. CoLSE is among the fastest learned models (\raisebox{0.1ex}{\scriptsize$\leq$}5 min), second only to LW-XGB (\raisebox{0.1ex}{\scriptsize$\leq$}5 min). Naru is slowest ($\sim$160 min). DeepDB spikes at $\rho = 0.6$ ($\sim$30 min), mirroring its latency pattern. MSCN and LW-NN are in mid-range (20–30 min). As expected, traditional methods incur minimal training: MHIST completes in under 10 minutes, while THIST and SRS build within seconds.

\begin{figure*}[htbp]
    % \centering
    \begin{subfigure}[b]{0.4\linewidth}
        \centering
        \includegraphics[width=\linewidth]{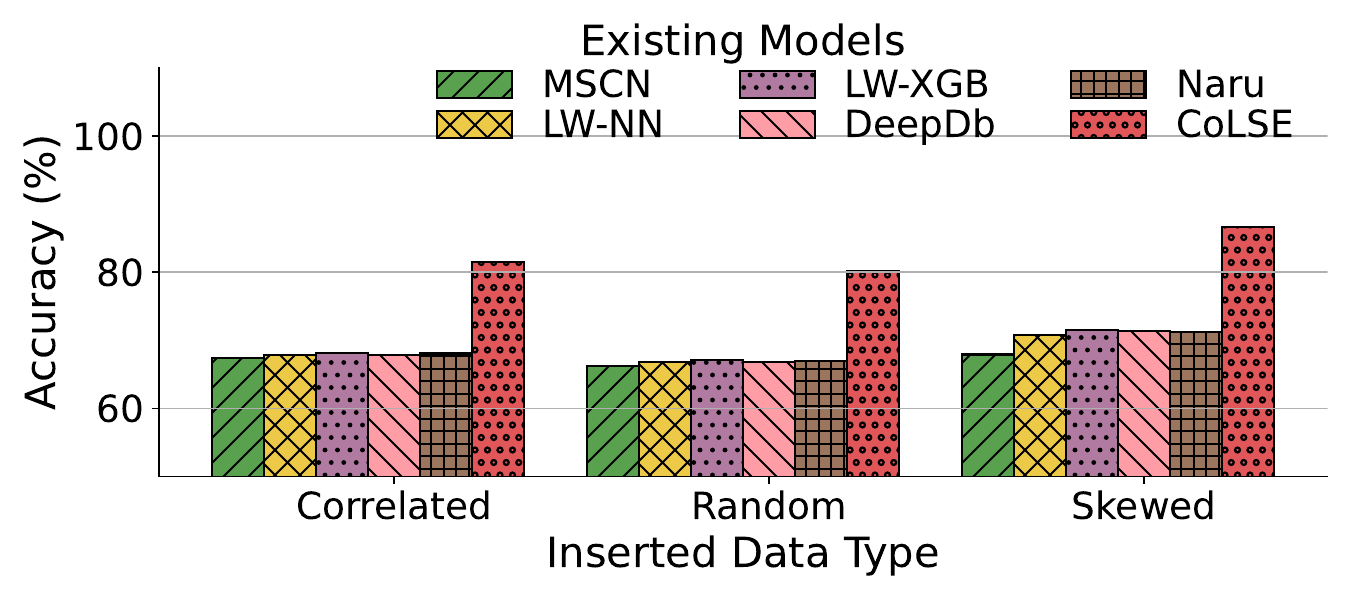}
        \vspace{-1.8em}
        \caption{Accuracy of models trained on past data}
        \label{fig:data_shift_accuracy_existing}
    \end{subfigure}
    % \hfill
    \begin{subfigure}[b]{0.4\linewidth}
        \centering
        \includegraphics[width=\linewidth]{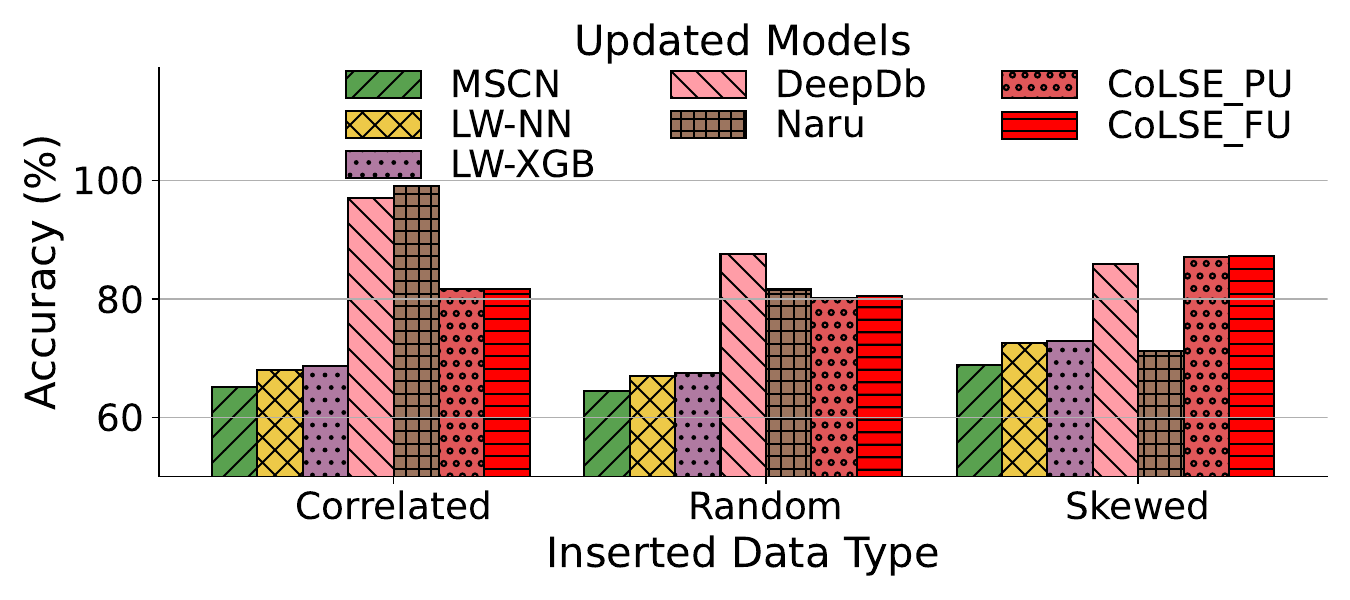}
        \vspace{-1.8em}
        \caption{Accuracy of models trained on new data}
        \label{fig:data_shift_accuracy_updated}
    \end{subfigure}
    \begin{subfigure}[b]{0.15\linewidth}
        % \centering
        \begin{minipage}{\linewidth}
            % \centering
            \resizebox{\linewidth}{!}{%
                \begin{tabular}{|c|c|}
                    \hline
                    Model & DMV \\
                    \hline
                    MSCN & 3.97 \\
                    LW-XGB & 1.43 \\
                    LW-NN & 6.83 \\
                    DeepDB & 0.19 \\
                    Naru & 1.29 \\
                    CoLSE\_PU & 0.05 \\
                    CoLSE\_FU & 1.55 \\
                    \hline
                \end{tabular}
                }
        \end{minipage}
        \caption{Updating times of models (minutes)}
        \label{fig:update_time_table}
    \end{subfigure}
    \vspace{-0.3em}
    \caption{Performance of models under different conditions of data updates on DMV data}
    \label{fig:dataupdates}
    % \vspace{-1.5em}
\end{figure*}

\subsubsection{Model Size Comparison}
% CoLSE yields the smallest model (around 2 MB), supporting deployment in memory-constrained environments. In contrast, Naru produces the largest (around 67 MB), owing to its deep autoregressive architecture.
% DeepDb shows a non-monotonic trend: starting at \raisebox{0.25ex}{\texttildelow}2 MB (correlation 0.2), peaking at \raisebox{0.25ex}{\texttildelow}17 MB (0.6), and dropping to \raisebox{0.25ex}{\texttildelow}8 MB (0.8), mirroring its latency and training time patterns.
% Among query-driven models, MSCN and LW-NN remain stable between 10–13 MB, while LW-XGB has a moderately larger footprint (\raisebox{0.25ex}{\texttildelow}25 MB), consistent with its ensemble structure.
CoLSE is most compact ($\sim$2 MB), enabling memory-constrained deployment. Naru is largest ($\sim$67 MB) due to its deep autoregressive design. DeepDB varies non-monotonically, $\sim$2 MB at $\rho = 0.2$, peaking $\sim$17 MB at $\rho = 0.6$, then $\sim$8 MB at $\rho = 0.8$, mirroring its latency/training trends. Among query-driven models, MSCN and LW-NN are stable between 10–13 MB, while LW-XGB is moderately larger ($\sim$25 MB) owing to its ensemble model.

\vspace{-0.5em}
\subsection{Impact of Data Updates}
\label{subsec:data_update}
Fig.~\ref{fig:dataupdates} illustrates how state-of-the-art models respond to different types of data updates before and after retraining, along with their respective update times. Each model is retrained following their original implementations. Specifically, MSCN, LW-XGB, and LW-NN are retrained using newly generated workloads comprising 10,000, 8,000, and 16,000 queries, respectively. Naru is updated by performing one additional training epoch, while DeepDB is incrementally updated by inserting a small sample (1\%) of the newly appended data into its tree-based model. To handle data updates, CoLSE can be retrained in two steps: 1) updating the marginal CDF distributions along with the dependency parameters needed for copula calculations-referred to as ``CoLSE\_PU'' (Partially Updated) and 2) additionally retraining the error compensation network-referred to as ``CoLSE\_FU'' (Fully Updated).

The Fig.~\ref{fig:data_shift_accuracy_existing} illustrates that all models experience some performance degradation following the insertion of new data. However, CoLSE exhibits the least deterioration, demonstrating greater robustness compared to other baselines across various types of data updates. After retraining (Fig.~\ref{fig:data_shift_accuracy_updated}), all models show improved accuracy, with data-driven models benefiting the most. While CoLSE does not consistently outperform all models post-retraining, it remains the most effective—particularly when the updated data is skewed. Notably, the performance of CoLSE\_PU and CoLSE\_FU is nearly identical, suggesting that retraining the error compensation network is not always necessary.

The Fig.~\ref{fig:update_time_table} presents the model updating times. Query-driven models generally require longer update times due to the overhead of query generation. Among them, LW-XGB exhibits the shortest update time. When CoLSE is fully updated, its update time remains comparable to that of LW-XGB, which is 1.5 minutes. However, CoLSE\_PU achieves smallest updating time across all the models, which is 0.05 minutes.

\begin{figure}[h]
    % \centering
    \begin{subfigure}[b]{0.47\linewidth}
        \centering
        \includegraphics[width=\linewidth]{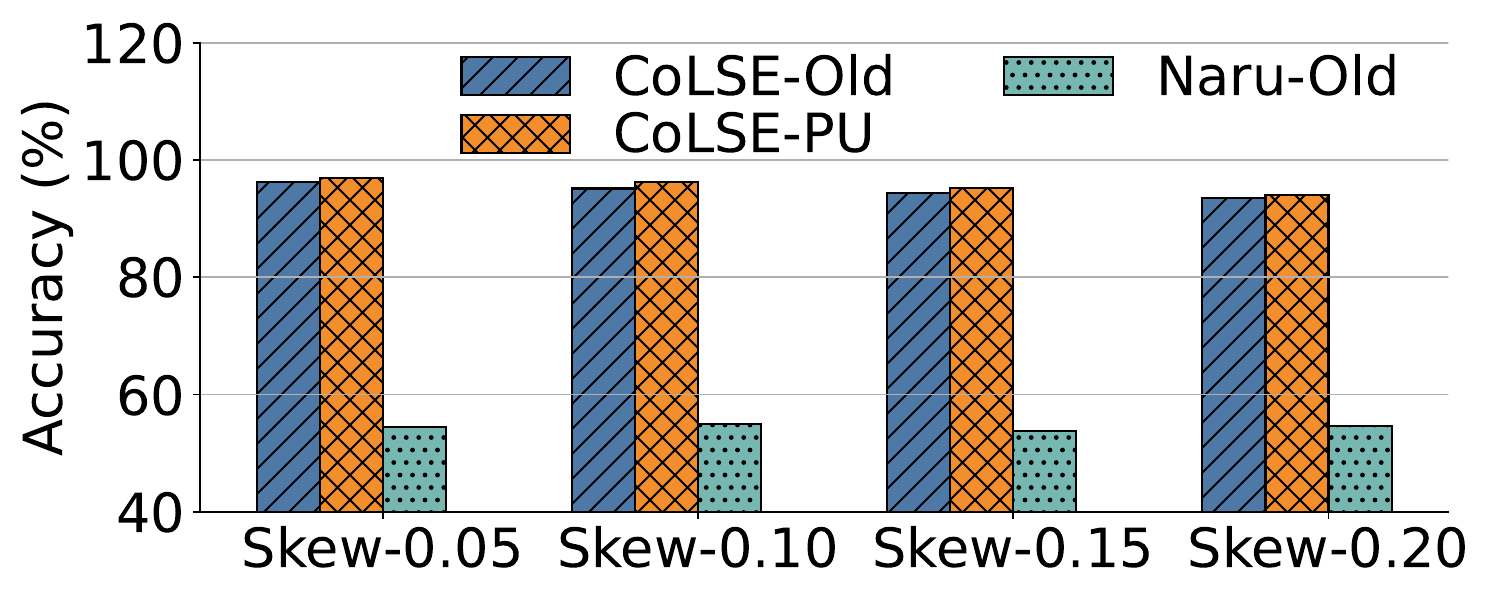}
        \vspace{-1.8em}
        % \caption{Accuracy of models trained on past data}
        % \label{fig:data_shift_accuracy_existing}
    \end{subfigure}
    % \hfill
    \begin{subfigure}[b]{0.47\linewidth}
        \centering
        \includegraphics[width=\linewidth]{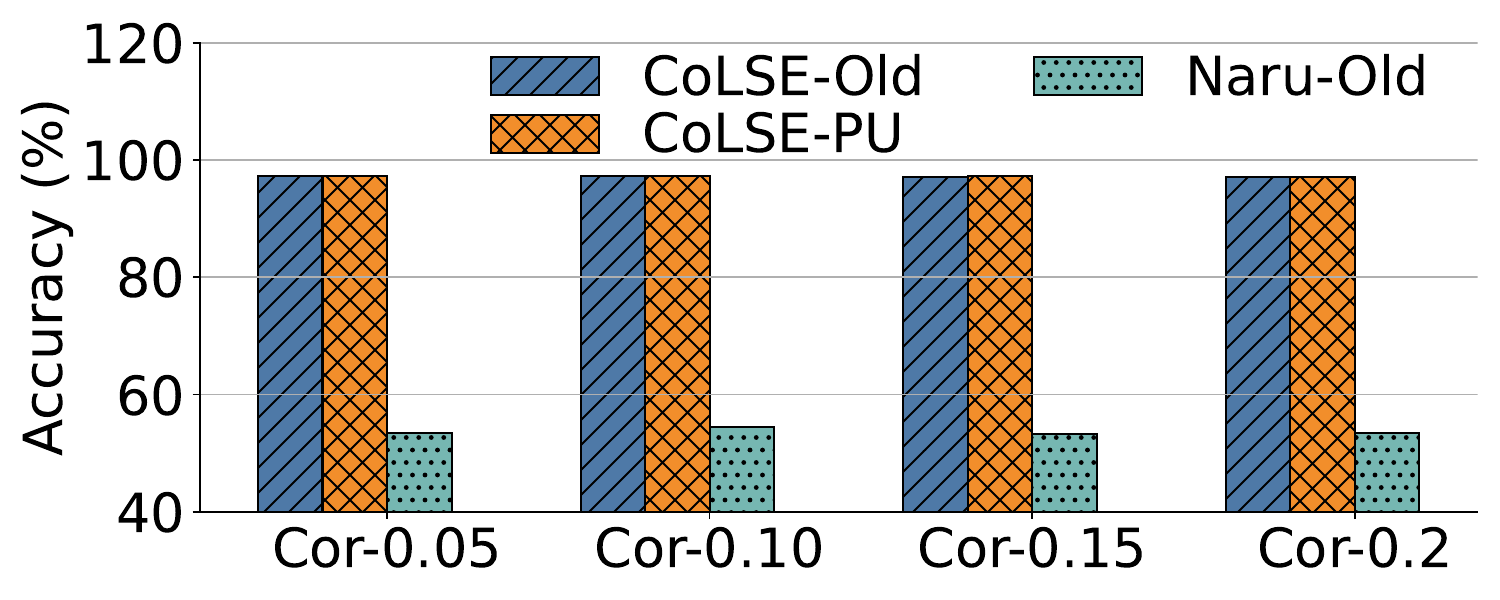}
        \vspace{-1.8em}
        % \caption{Accuracy of models trained on new data}
        % \label{fig:data_shift_accuracy_updated}
    \end{subfigure}
    \caption{Performance of CoLSE vs. Naru on DMV under continual updates}
    \label{fig:updateheavy}
    \vspace{-1.5em}
\end{figure}

We evaluate robustness on DMV in update-heavy settings by inserting 5\% new tuples after each query batch for four rounds (20\% total), with skewed or correlated drift (Fig.~\ref{fig:updateheavy}). Here, CoLSE-old and Naru-old refer to existing models without retraining. Overall, CoLSE-Old remains accurate, while CoLSE-PU provides small, consistent gains over the existing model. In contrast, Naru degrades as drift accumulates, indicating that CoLSE is robust under continual updates.

\subsection{Impact of Workload Shifts}
\label{subsec:workload_shift}

\begin{figure}
    \centering
    \includegraphics[width=0.84\linewidth]{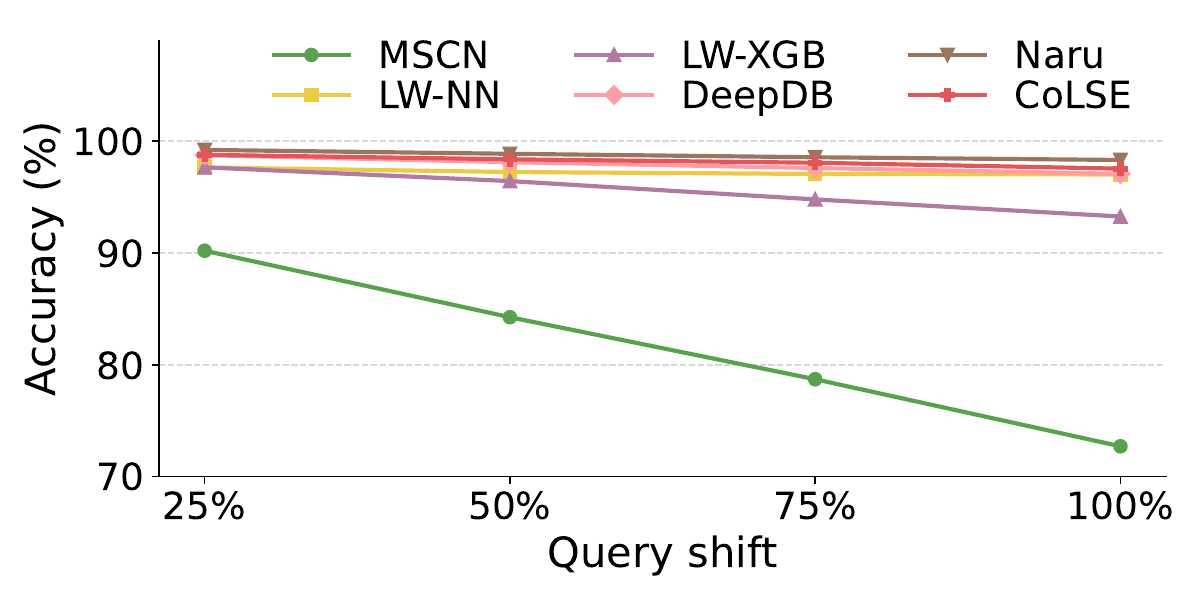}
    \vspace{-1em}
    \caption{Performance of models with dynamic workload distribution shifts on the DMV dataset}
    \label{fig:queryshift}
    % \vspace{-1.5em}
\end{figure}

Fig.~\ref{fig:queryshift} illustrates how state-of-the-art models respond to shifts in query distributions.  
We simulate workload shifts by incrementally replacing 25\% of the original workload with queries drawn from a different distribution at each step.

As the shift ratio increases,  
CoLSE demonstrates robustness comparable to data-driven methods, consistently outperforming all query-driven baselines. This indicates that CoLSE not only captures complex data correlations but also generalizes well to previously unseen query patterns.  
As expected, data-driven models exhibit greater resilience to distributional drift due to their explicit modeling of the joint data distribution.  
In contrast, query-driven methods tend to overfit to the observed workload, resulting in noticeable performance degradation as the distribution shift increases.  
For example, MSCN suffers a substantial drop in accuracy when the original workload is fully replaced.

These findings highlight the limitations of purely query-driven estimators and underscore CoLSE’s effectiveness in maintaining stable accuracy under evolving workloads.
\vspace{-0.5em}
% As anticipated, data‑driven approaches exhibit greater resilience to these shifts than query‑driven methods, with performance degrading as the magnitude of the distribution shift increases. Interestingly, CoLSE shows a similar performance to data-driven methods outperforming all the query‑driven techniques.

\begin{figure*}[htbp]
     \centering
    \begin{subfigure}[b]{0.39\linewidth}
        \centering
        \includegraphics[width=\linewidth]{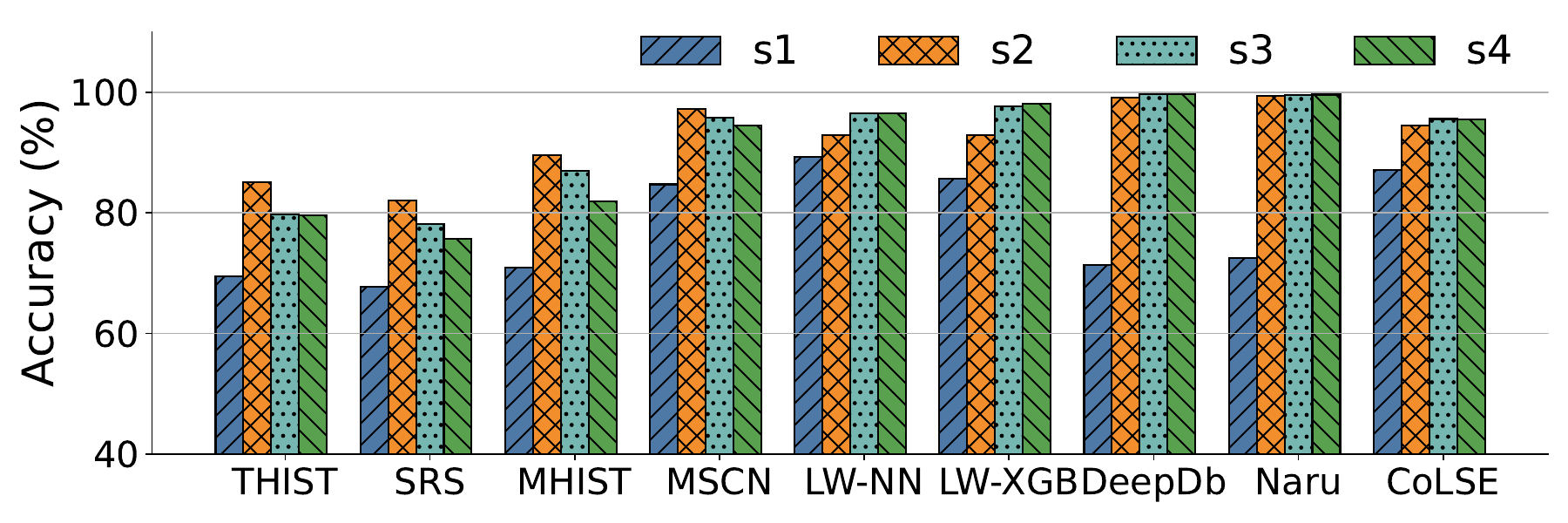}
        \vspace{-1.8em}
        \caption{Accuracy}
        \label{fig:skew_accuracy}
    \end{subfigure}
    % \hfill
    \begin{subfigure}[b]{0.39\linewidth}
        \centering
        \includegraphics[width=\linewidth]{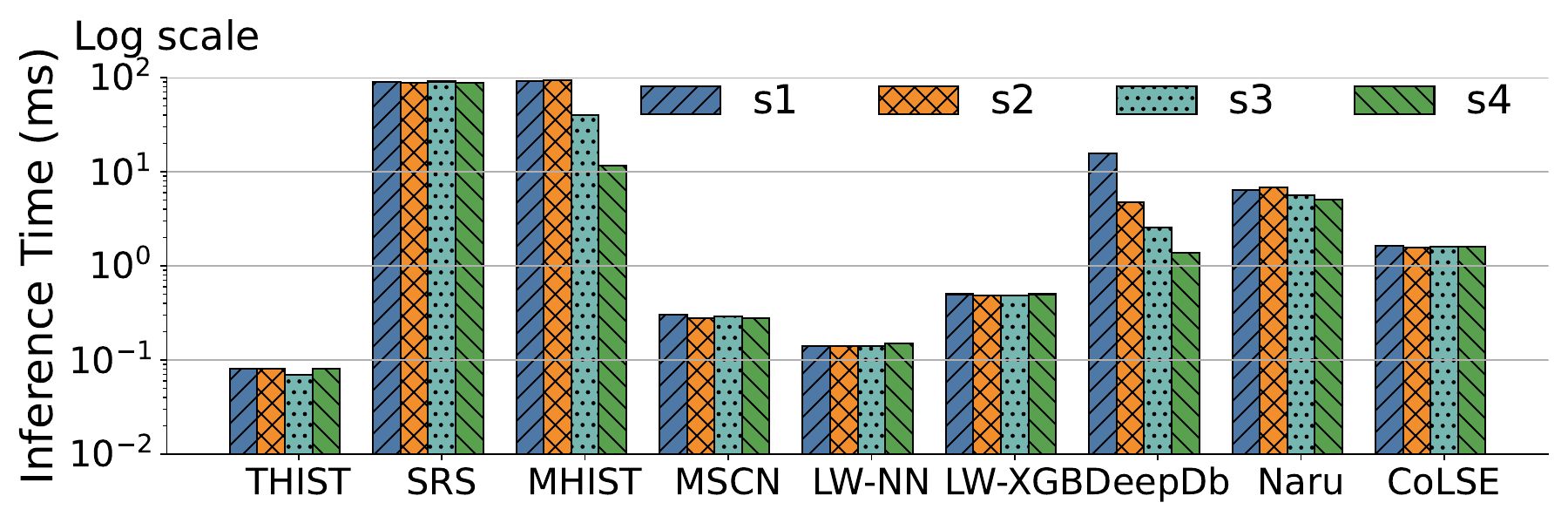}
        \vspace{-1.8em}
        \caption{Inference time (ms)}
        \label{fig:skew_inference_time}
    \end{subfigure}
    \\
    \begin{subfigure}[b]{0.39\linewidth}
        \centering
        \includegraphics[width=\linewidth]{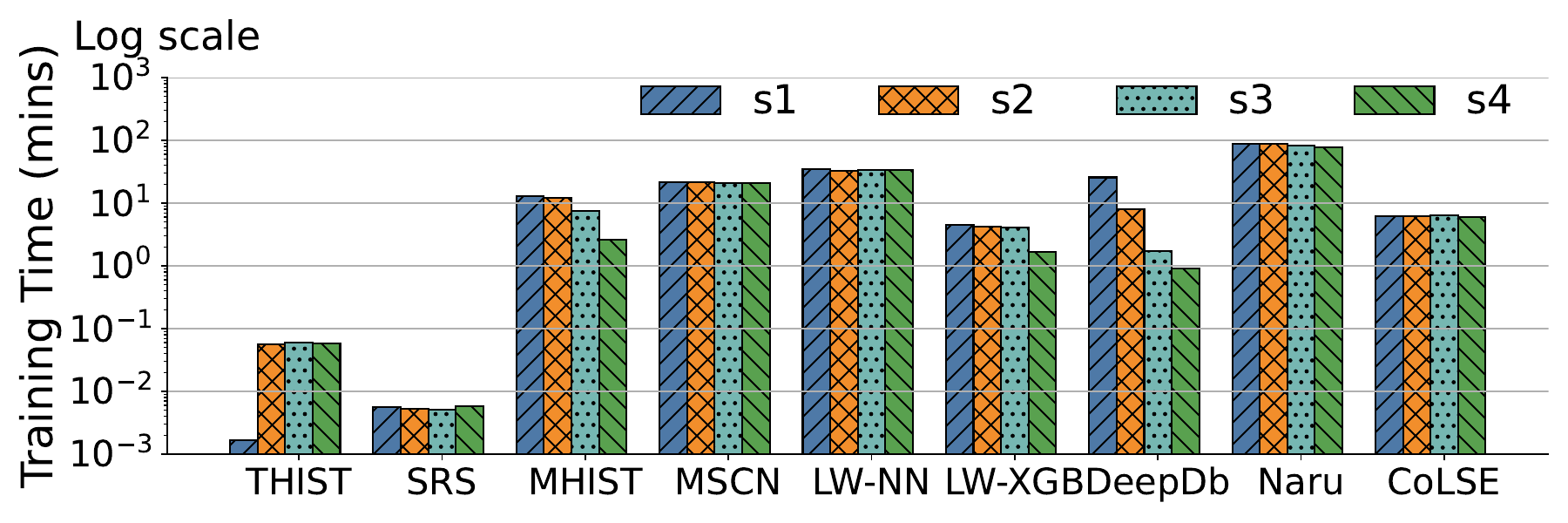}
        \vspace{-1.8em}
        \caption{Training time (min)}
        \label{fig:skew_training_time}
    \end{subfigure}
    % \hfill
    \begin{subfigure}[b]{0.39\linewidth}
        \centering
        \includegraphics[width=\linewidth]{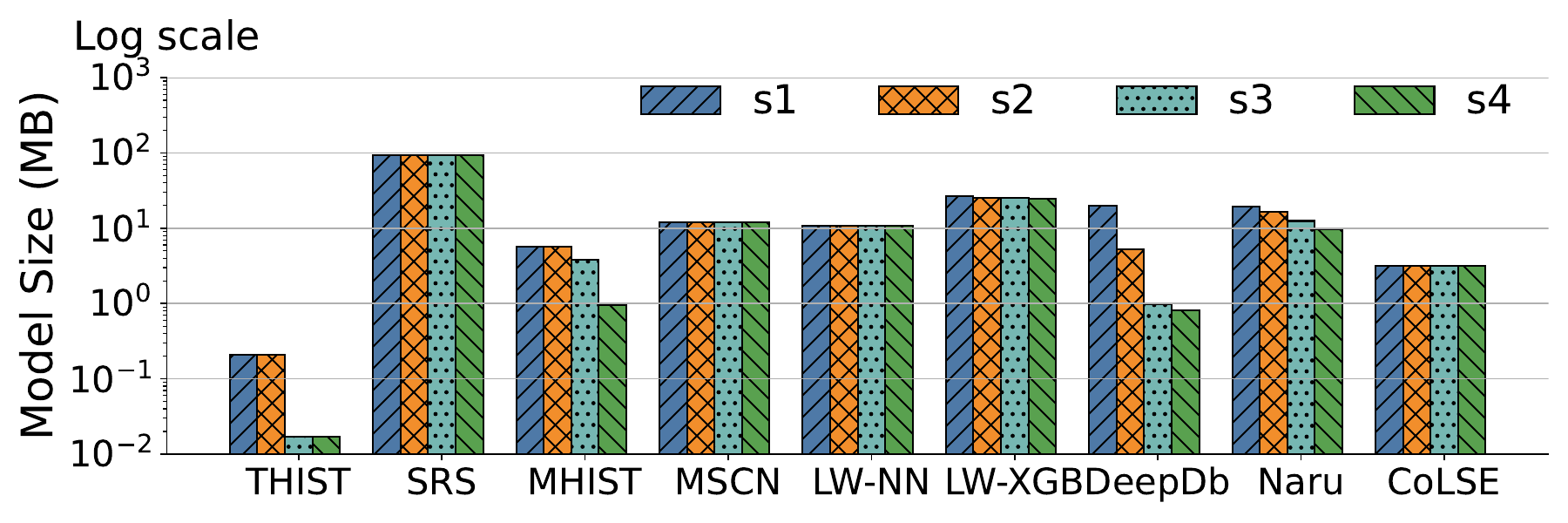}
        \vspace{-1.8em}
        \caption{Model size (MB)}
        \label{fig:skew_model_size}
    \end{subfigure}
    \vspace{-0.5em}
    \caption{Varying skew levels}
    \label{fig:skew}
    % \vspace{-0.5em}
\end{figure*}
\subsection{Impact of Data Skew}
\label{subsec:skew}
\subsubsection{Accuracy Comparison}
CoLSE maintains strong and consistent accuracy across varying levels of data skew. While it does not always outperform every baseline in absolute accuracy, its performance remains competitive and stable, even under challenging low-skew scenarios, making it a reliable choice across diverse distributions.

As skew increases, most learned models—especially data-driven ones like DeepDb and Naru—show improved accuracy, benefiting from their ability to learn structured patterns in non-uniform data. For instance, DeepDb improves from roughly 77\% to 83.5\% as skew increases. Conversely, traditional methods degrade under skew due to their limited modeling capacity.  
Notably, MSCN, a query-driven model, fails to capitalize on skewness, likely due to its reliance on sampling-based features. Meanwhile, LW-XGB, though highly accurate at low skew, shows reduced robustness at higher skew levels.

Data skewness clearly has a notable impact on estimation performance. Interestingly, all models, including learned ones, perform worst on the dataset with minimal skew. This dataset contains a high number of distinct values with low true cardinalities, leading the optimizer to favor index scans, which are highly sensitive to estimation errors. Even minor inaccuracies in such settings can result in suboptimal plan choices and significant deviations.

\subsubsection{Inference Latency Comparison}
% As shown in Fig.~\ref{fig:skew_inference_time}, CoLSE positions itself between query-driven and data-driven methods, with inference latency closer to query-driven approaches while maintaining consistent performance across different skew levels (less than 2 ms).  
% In contrast, query-driven models show the lowest and most consistent inference times (near or below 0.5 ms). DeepDb exhibits the highest inference latency (16 ms) at the lowest skew level (s1), but its inference time steadily decreases as skew increases, dropping below 2 ms at s4. This trend is primarily due to a reduction in unique values, which simplifies the data distribution and allows models to converge to smaller, more efficient representations. Naru follows a similar pattern but with moderately higher latency (ranging from 7 ms to 5 ms).  
% Traditional methods generally exhibit higher inference times compared to learned approaches.
As shown in Fig.~\ref{fig:skew_inference_time}, CoLSE remains under 2 ms across skew levels, positioning it near query-driven methods, which are most consistent at \raisebox{0.1ex}{\scriptsize$\leq$}0.5 ms. DeepDB is highest at low skew ($\sim$16 ms at s1) but drops below 2 ms by s4, as fewer distinct values simplifies the distribution and yields smaller, more efficient models. Naru follows the same trend with moderately higher latency ($\sim$7 to 5 ms). Traditional methods generally exhibit higher inference times compared to learned approaches.

% \begin{figure*}[htbp]
%     \centering
%     \includegraphics[width=0.95\linewidth]{figures/model_size.png}
%     \caption{(a) No. of optimal query plans(\%) (b) Inference time(ms) (c) Training time(min) and (d) Model size(MB) across learned approaches at varying dataset sizes}
%     \label{fig:datasize}
% \end{figure*}

\begin{figure*}[htbp]
     \centering
    \begin{subfigure}[b]{0.39\linewidth}
        \centering
        \includegraphics[width=\linewidth]{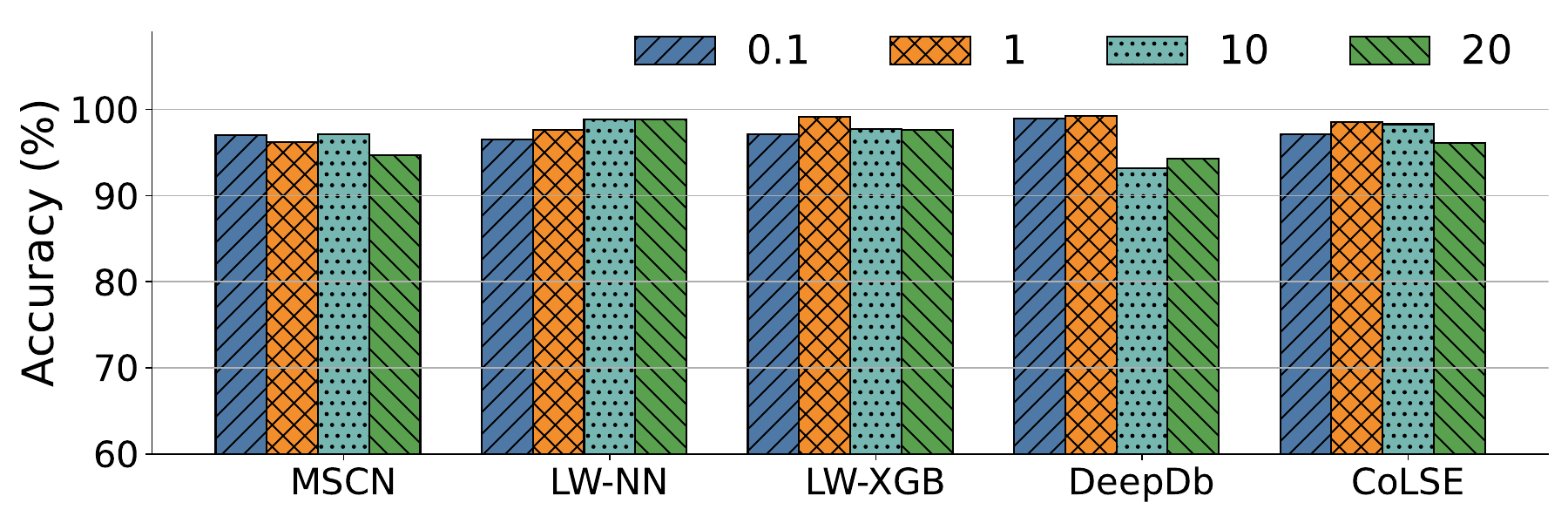}
        \vspace{-1.8em}
        \caption{Accuracy}
        \label{fig:data_size_accuracy}
        
    \end{subfigure}
    % \hfill
    \begin{subfigure}[b]{0.39\linewidth}
        \centering
        \includegraphics[width=\linewidth]{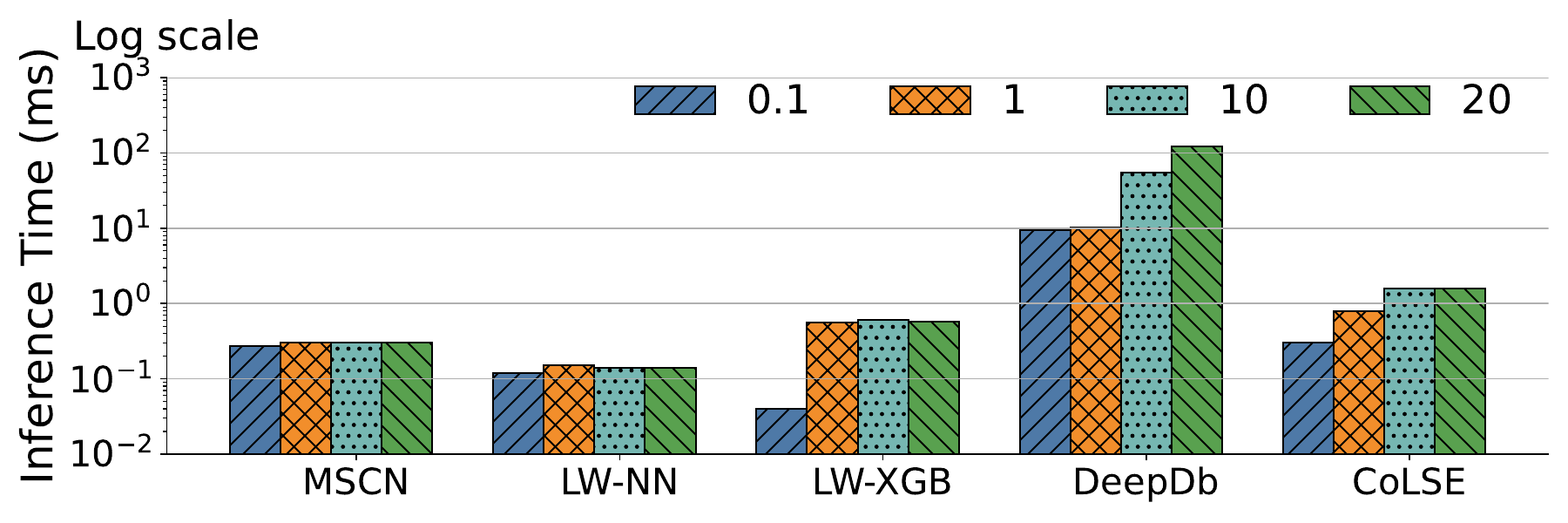}
        \vspace{-1.8em}
        \caption{Inference time (ms)}
        \label{fig:data_size_inference_time}
        
    \end{subfigure}
    \\
    \begin{subfigure}[b]{0.39\linewidth}
        \centering
        \includegraphics[width=\linewidth]{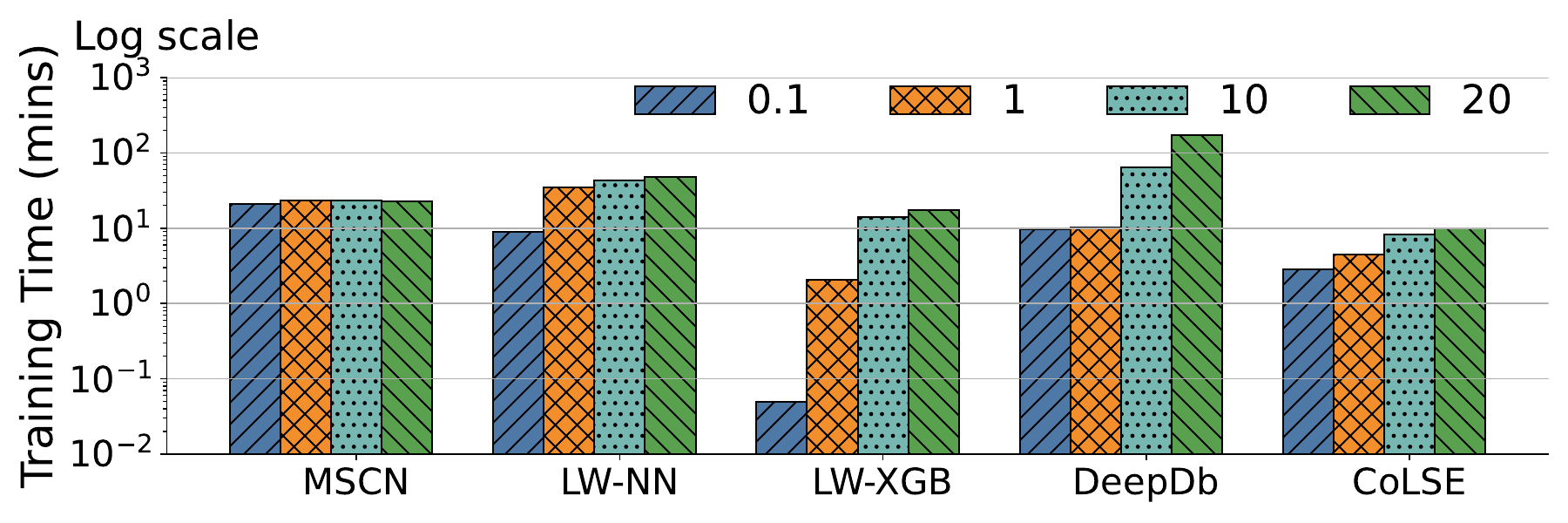}
        \vspace{-1.8em}
        \caption{Training time (min)}
        \label{fig:data_size_training_time}
        
    \end{subfigure}
    % \hfill
    \begin{subfigure}[b]{0.39\linewidth}
        \centering
        \includegraphics[width=\linewidth]{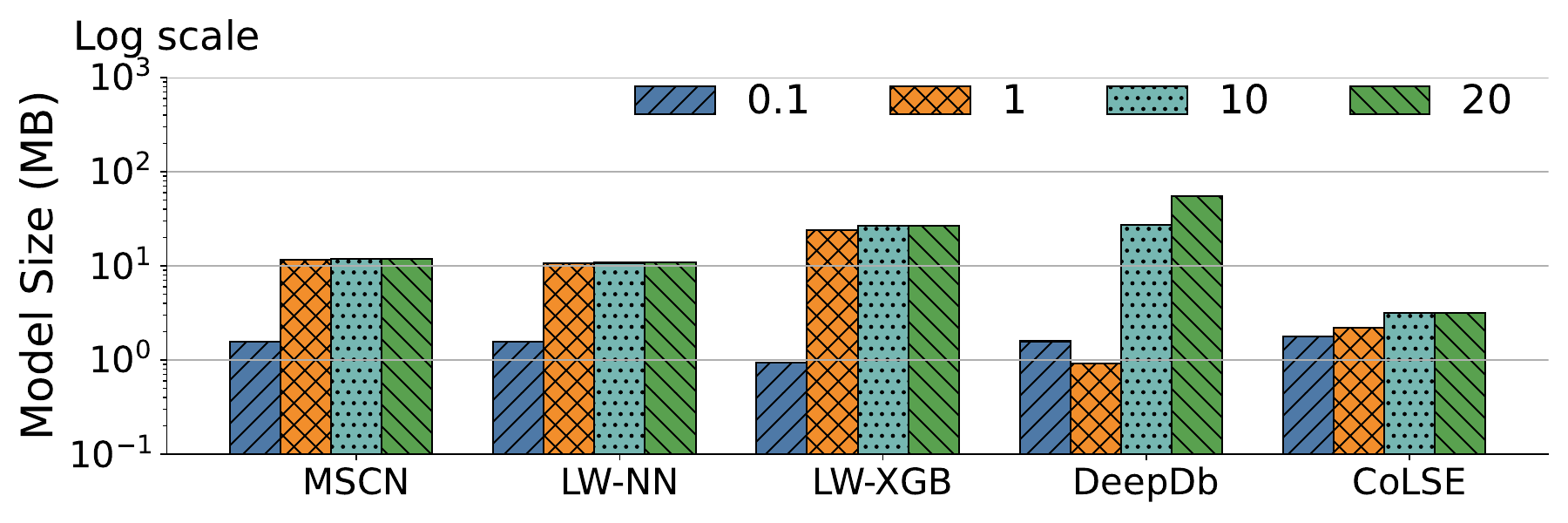}
        \vspace{-1.8em}
        \caption{Model size (MB)}
        \label{fig:data_size_model_size}
    \end{subfigure}
    \vspace{-1em}
    \caption{Varying dataset sizes}
    \label{fig:datasize}
    % \vspace{-1em}
\end{figure*}

\subsubsection{Training Time Comparison}
% Fig.~\ref{fig:skew_training_time} compares training times across skew levels.  
% CoLSE demonstrates low and consistent training times (around 6 minutes), making it one of the most training-efficient learned models.  
% Traditional methods show lower training times overall, as expected. Among the learned models, Naru has the longest training duration, decreasing from approximately 90 minutes at s1 to 75 minutes at s4. DeepDb follows a similar pattern but with significantly lower values, around 28 min at s1, dropping to less than 5 min at s4. LW-NN, LW-XGB, and MSCN exhibit moderate and relatively stable training times across all skew levels (around 20–35 minutes).
%\lankadinee{Fig.~\ref{fig:skew_training_time} shows that CoLSE trains rapidly and consistently (\raisebox{0.25ex}{\texttildelow}6 min across skews), making it one of the most training-efficient learned models. Traditional methods are faster overall, as expected. Among learned baselines, Naru is longest (\raisebox{0.25ex}{\texttildelow}90 to 75 min from s1 to s4), while DeepDB drops sharply (\raisebox{0.25ex}{\texttildelow}28 min at s1 to \raisebox{0.1ex}{\scriptsize$\leq$}5 min at s4). LW-NN, LW-XGB, and MSCN are moderate and relatively stable (\raisebox{0.25ex}{\texttildelow}20–35 min).}

Fig.~\ref{fig:skew_training_time} shows that CoLSE trains quickly and consistently ($\sim$6 min across skews), making it one of the most efficient learned models. Expectedly, traditional methods are faster overall. Among learned baselines, Naru is the slowest ($\sim$90 to 75 min from s1 to s4), while DeepDB drops sharply (28 min to \raisebox{0.1ex}{\scriptsize$\leq$}5 min). LW-NN, LW-XGB, and MSCN show moderate, stable times (20–35 min).

\subsubsection{Model Size Comparison}
% As shown in Fig.~\ref{fig:skew_model_size}, CoLSE achieves the smallest model size among all learned methods, maintaining a consistent footprint of less than 4 MB across all skew scenarios, highlighting its scalability.  
% Conversely, LW-XGB exhibits the largest model size, which remains consistently high across skew levels. DeepDb and Naru show a decreasing trend in model size as skew increases, consistent with their training time patterns.
As shown in Fig.~\ref{fig:skew_model_size}, CoLSE is the most compact, remaining \raisebox{0.1ex}{\scriptsize$\leq$}4 MB across all skew levels. Conversely, LW-XGB is largest and stays high regardless of skew. DeepDB and Naru shrink as skew increases, mirroring their training-time trends.

\vspace{-0.5em}
\subsection{Impact of Dataset Size}
\label{subsec:dataset_size}
% Experimental results in Section~\ref{subsec:sota} show that larger datasets dramatically increase training time, inference time, and model size for learned methods, despite these methods outperforming traditional approaches in terms of accuracy. 
Section~\ref{subsec:sota} shows that larger datasets inflate training/inference time and model size for learned methods (though they remain more accurate than traditional baselines).
Hence, this evaluation focuses solely on learned models to assess their performance on large-scale datasets. Naru is omitted for 10 and 20 GB datasets since one epoch on 10 GB took $\sim$22 h, indicating poor scalability beyond 10 GB.

% Results for Naru are not reported for the 10 GB and 20 GB datasets, as training for just one epoch on the 10 GB dataset took approximately 22 hours. This highlights a significant scalability limitation, indicating that Naru is not suitable for handling very large datasets (above 10 GB) within practical time constraints. 

\subsubsection{Accuracy Comparison}
%Fig.~\ref{fig:data_size_accuracy} shows that even for larger datasets, CoLSE achieves accuracy comparable to other learned approaches, consistently exceeding 95\%. In particular, on the 10 GB and 20 GB datasets, it outperforms MSCN and DeepDb.
Fig.~\ref{fig:data_size_accuracy} shows that CoLSE maintains over 95\% accuracy even on larger datasets, matching or surpassing other learned methods. On the 10 GB and 20 GB datasets, it outperforms MSCN and DeepDb.

\subsubsection{Inference Latency Comparison}
% Fig.~\ref{fig:data_size_inference_time} compares inference times. Even with larger datasets, query-driven models achieve near-zero inference latency. In contrast, DeepDb shows a dramatic increase in inference time as the dataset size grows—starting modestly for smaller datasets but escalating sharply to approximately 125 ms at 20 GB. This spike reflects its high computational complexity during prediction, making it less practical for large-scale applications.  
% Overall, CoLSE offers a favorable trade-off, delivering significantly lower latency than data-driven models while remaining competitive with query-driven methods.
Fig.~\ref{fig:data_size_inference_time} compares inference times. Even with larger datasets, query-driven models remain near-zero. DeepDB escalates with size, reaching $\sim$125 ms at 20 GB, reflecting high computational complexity during predication and limiting practicality. CoLSE offers a favorable trade-off—much lower latency than data-driven methods while remaining competitive with query-driven ones.

\subsubsection{Training Time Comparison}
% As the dataset size increases, training times tend to rise across all models except for MSCN (see Fig.~\ref{fig:data_size_training_time}, which shows consistent training times of around 20 minutes across all datasets. DeepDb exhibits a sharp escalation, peaking at approximately 175 minutes (nearly 3 hours) for the 20 GB dataset. In contrast, LW-NN shows a gradual increase, starting at approximately 10 minutes for the smallest dataset and reaching nearly 50 minutes at 20 GB. LW-XGB and CoLSE follow a similar trend, though more moderate, with training times increasing from a few minutes to around 20 minutes and 10 minutes respectively, as dataset size grows.
Training time generally increases with dataset size, except for MSCN, which remains roughly constant at $\sim$20 minutes across all scales (Fig.~\ref{fig:data_size_training_time}). DeepDB scales poorly, rising sharply to $\sim$175 minutes ($\sim$3 hours) at 20 GB. In contrast, LW-NN grows gradually from $\sim$10 to $\sim$50 minutes. LW-XGB and CoLSE also increase with size but remain moderate, reaching $\sim$20 minutes and $\sim$10 minutes, respectively, at 20 GB.

\subsubsection{Model Size Comparison}
% Fig.~\ref{fig:data_size_model_size} shows that as dataset size increases, model sizes generally grow across all methods. CoLSE consistently maintains the smallest and most stable footprint, remaining around 2–3 MB. In contrast, DeepDb grows significantly, exceeding 55 MB at 20 GB. Among query-driven methods, LW-XGB sees the most notable increase, reaching about 27 MB in the 10–20 GB range, despite having the lowest training and inference times.
As Fig.~\ref{fig:data_size_model_size} depicts, model sizes generally grow with dataset size. CoLSE remains smallest and stable ($\sim$2–3 MB), DeepDB expands steeply (\raisebox{0.1ex}{\scriptsize\textgreater}55 MB at 20 GB), and among query-driven methods LW-XGB grows most ($\sim$27 MB at 10–20 GB) despite its speed advantages.

\begin{table}[tb]
\caption{Performance of different copula types on different correlation structures}
\vspace{-1em}
\begin{center}
\begin{tabular}{|l|c|c|c|}
\hline
\textbf{Column groups} & \textbf{Frank} & \textbf{Clayton} & \textbf{Gumbel} \\
\hline
[0, 1, 2]    & 1.03E-03  & 1.18E-04 & \textbf{8.28E-05} \\
\hline
[2, 3, 4]   & 8.82E-04  & 8.29E-04 & \textbf{7.68E-04} \\
\hline
[5, 6, 7]  & 9.35E-04 & 1.40E-04 & \textbf{1.16E-04} \\
\hline
[8, 9, 10]  & 6.47E-04  & 3.51E-04 & \textbf{3.26E-04} \\
\hline
[4, 5, 6]    & 0.001  & 0.001 & \textbf{0.001} \\
\hline
[7, 8, 9]   & 0.001  & 0.001 & \textbf{0.001} \\
\hline
\end{tabular}
\label{tab:copulaperformance}
\end{center}
 \vspace{-1em}
\end{table}

\begin{figure*}[htbp]
    \centering
    % === Figures block ===
    \begin{minipage}[t]{0.7\linewidth}
        \centering
        \begin{subfigure}[b]{0.28\linewidth}
            \centering
            \includegraphics[width=\linewidth]{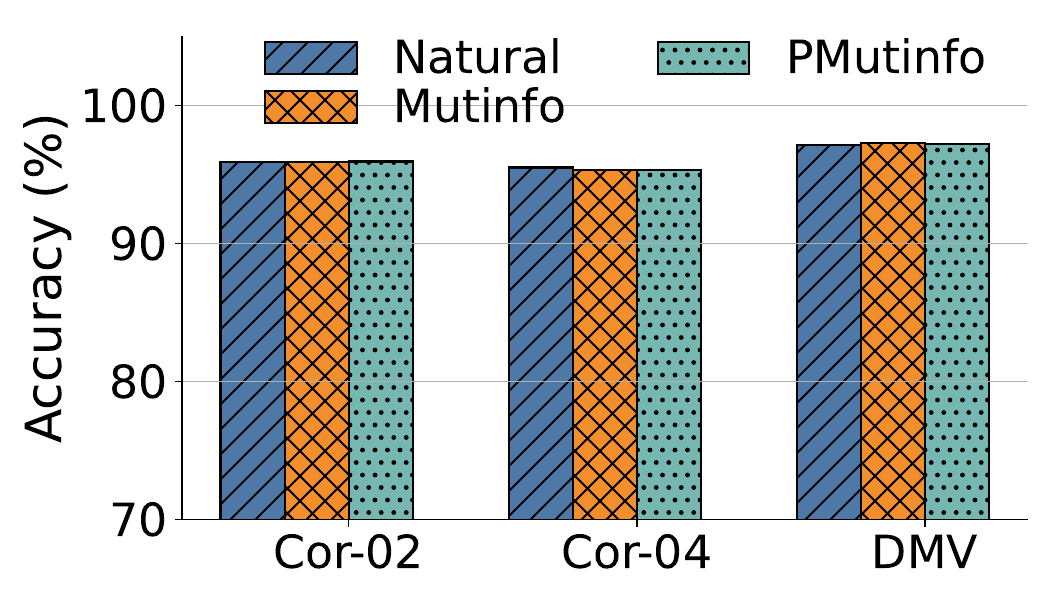}
            \vspace{-1.8em}
            % \caption{\revision{Attribute ordering}}
            \caption{}
            \label{fig:attrorder}
        \end{subfigure}
        \begin{subfigure}[b]{0.28\linewidth}
            \centering
            \includegraphics[width=\linewidth]{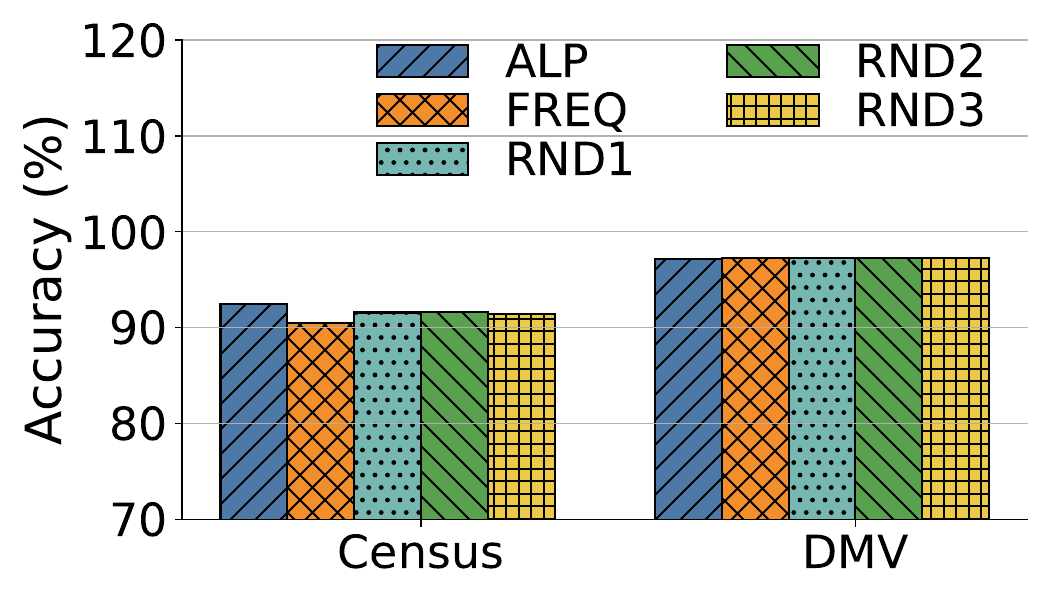}
            \vspace{-1.8em}
            % \caption{\revision{Categorical encoding}}
            \caption{}
            \label{fig:catlabelencoding}
        \end{subfigure}
        \begin{subfigure}[b]{0.4\linewidth}
            \centering
            \includegraphics[width=\linewidth]{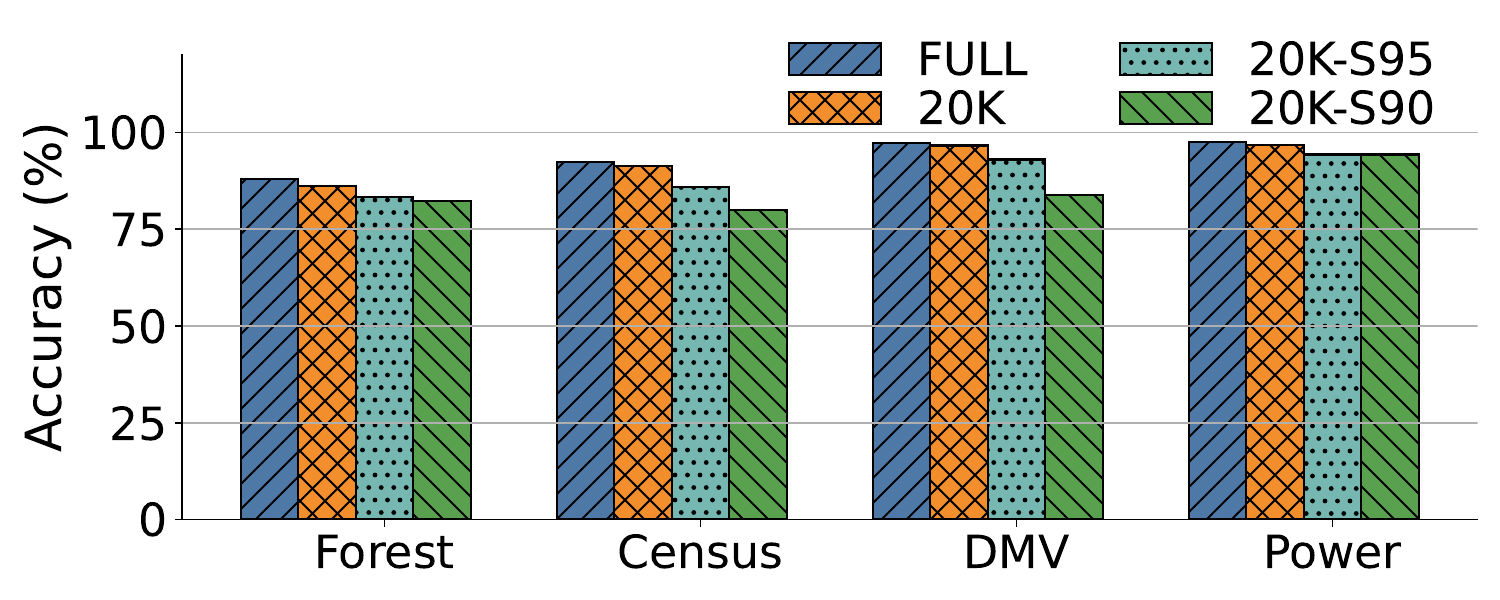}
            \vspace{-1.8em}
            % \caption{\revision{Sparse workloads}}
            \caption{}
            \label{fig:sparsewkldexp}
        \end{subfigure}
        \vspace{-0.5em}
        \caption{Sensitivity study: Impact of (a) attribute ordering and (b) label encoding\\ variants on JPE performance, and (c) Impact of sparse training workloads on ECN}
        \label{fig:sensitivity_study}
    \end{minipage}%
    \hfill
    % === Table block ===
    \begin{minipage}[t]{0.3\linewidth}
        \vspace{-65pt} % ensures alignment at top
        \centering
        \resizebox{\linewidth}{!}{%
            \begin{tabular}{|p{1cm}|p{0.9cm}|p{1.1cm}|p{1.1cm}|p{0.9cm}|}
                \hline
                Model & E2E Time (s) & Infer Time (ms) & Training Time (min) & Model size (MB)\\
                \hline
                ASM & 3.97 & 9.84 & 38 & 42 \\
                PRICE & 1.43 & 20.13 & 10 & 42 \\
                CoLSE & 6.83 & 1.25 & 05 & 28\\
                \hline
            \end{tabular}
        }
        \captionof{table}{Performance comparison on joins}
        \label{tab:joinresults}
    \end{minipage}
\end{figure*}

\vspace{-0.3em}
\subsection{Sensitivity Analysis}
\label{subsec:sensitivityanalysis}
\vspace{-0.2em}
Our model includes several parameters associated with its two components. We first focus on those related to the joint probability estimator. As explained in Section~\ref{sec:novelalgo}, the Gumbel copula is selected as the copula family for all pair-copula modeling. Table~\ref{tab:copulaperformance} supports this choice, as Gumbel consistently yields the lowest error across different correlation structures compared to other popular Archimedean copulas~\cite{genest1993statistical}, namely Frank and Clayton.

\textbf{Attribute Ordering in JPE.} We assess the impact of attribute order on JPE performance using three strategies from~\cite{yang2019deep}: MutInfo (maximizing mutual information with the selected set), PMutInfo (mutual information with only the most recent attribute), and the Natural ordering (schema order). Experiments on synthetic datasets (pairwise correlations 0.2 and 0.6) and the DMV dataset show only minor differences (see Fig.~\ref{fig:attrorder}). Interestingly, the natural ordering is also effective, likely reflecting a human bias to place key columns early, thereby reducing the uncertainty of subsequent attributes~\cite{yang2019deep}. Since MutInfo and PMutInfo are expensive and scale poorly, while Natural incurs no extra computation, we adopt Natural as the practical default.

%\revision{\info{R2:O3} \textbf{Categorical Attribute Encoding Schemes.} Constructing a CDF for categorical variables requires imposing an order, which is arbitrary for nominal attributes and thus a limitation~\cite{Mizuno2022SIS, Categorical_Distribution_2025}. To reduce bias from fixed numeric codes, our spline-based dequantization (Section~\ref{subsec:catvarhandling}) uses this order solely to define CDF steps, not as a direct embedding, thereby avoiding artificial ordinal distances~\cite{dunn1996randomized}. We evaluated the impact of ordering by comparing alphabetical, frequency-based, and randomly permuted sequences on two real-world datasets (Census and DMV). Results were consistent across all configurations, indicating that the imposed ordering has negligible effect on model performance. Given this, we adopt alphabetical ordering as the default for practicality and reproducibility.}

\textbf{Categorical Attribute Encoding Schemes.} Constructing a CDF for categorical variables requires an imposed order, which is arbitrary for nominal attributes --a known limitation~\cite{Mizuno2022SIS, Categorical_Distribution_2025}. To mitigate bias from fixed numeric codes, our spline-based dequantization (Section~\ref{subsec:catvarhandling}) uses the order only to define CDF steps, not for direct embedding, thus avoiding artificial ordinal distances~\cite{dunn1996randomized}. We assessed the effect of ordering using alphabetical, frequency-based, and random permutations on two real-world datasets (Census and DMV). Results were consistent, showing negligible impact on performance. For practicality and reproducibility, we thus default to alphabetical ordering.

\begin{figure*}[htbp]
    \centering
    \includegraphics[width=0.94\linewidth]{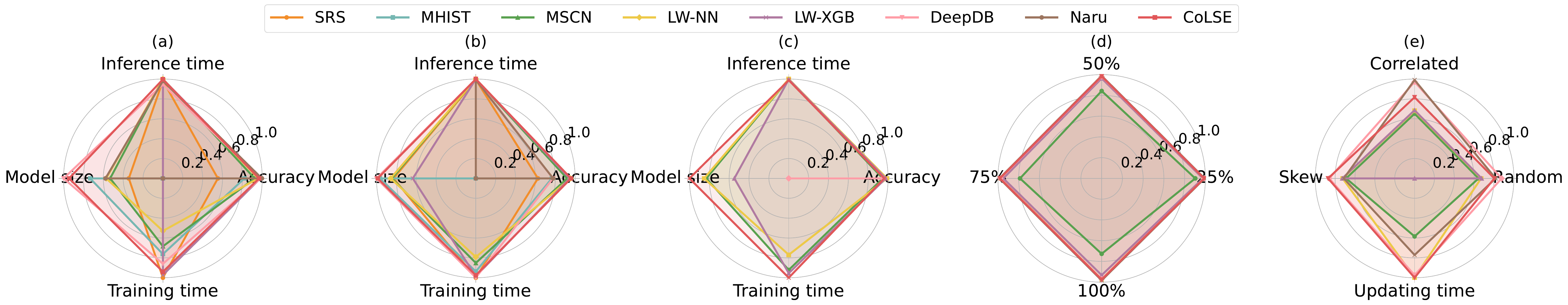}
    \caption{Summary of experimental evaluation: (a) on a real-world dataset (DMV), (b) on a correlated dataset with a pairwise correlation coefficient of 0.2, (c) on a large-scale dataset of size 20 GB, (d) under query workload shift (using DMV) and (e) with different data updates applied to DMV dataset}
    % \vspace{-1em}
    \label{fig:spider}
\end{figure*}

%\revision{\info{R3:O2} \textbf{Error bounds for JPE.} We estimated empirical error bounds for JPE using a two-stage procedure: bootstrapped CIs of log-scale multiplicative errors within each dataset, followed by a random-effects meta-analysis across datasets. The resulting error bounds are $1.09×–2.89×$, indicating that, on average, the model’s estimates can be off by between 1.1- and 2.9-fold.}
\textbf{Error Bounds for JPE.} We estimate empirical error bounds for JPE using a two-stage approach: bootstrapped confidence intervals of log-scale multiplicative errors per dataset, followed by a random-effects meta-analysis across datasets. The resulting bounds--$1.09×$ to $2.89×$--indicate that, on average, estimates may vary by 1.1× to 2.9× from true values.

%For the error compensation network, we evaluated how validation loss varied across several network architectures with different hidden layer sizes: $512\_256$, $256\_256\_256$, $256\_256\_128\_128$, $256\_256\_64$, and $256\_256\_128\_64$.  
% The Fig.~\ref{fig:sensitivityanalysis}(c) illustrates the average of the validation losses obtained for Power and Forest datasets under each configuration. 
%Among these, the $256\_256\_128\_64$ configuration achieved the lowest average validation loss when trained with a batch size of 32, a learning rate of 0.001, and 25 epochs. As a result, this architecture was adopted for all datasets without further hyperparameter tuning, offering a strong balance between predictive accuracy, model size, and training time. This choice is further supported by the nature of the task: the error compensation network corrects errors from the joint probability estimator—a problem typically involving simpler patterns and lower variance than modeling the full joint distribution.
We evaluated several architectures for the error compensation network, $512\_256$, $256\_256\_256$, $256\_256\_128\_128$, $256\_256\_64$, and $256\_256\_128\_64$, by comparing validation loss on the Power and Forest datasets. The $256\_256\_128\_64$ configuration yielded the lowest average loss using a batch size of 32, learning rate of 0.001, and 25 training epochs. We selected this architecture for all datasets without further tuning, as it balances accuracy, model size, and training time. This is also aligned with the task's nature: correcting residual errors from the joint estimator generally requires less model complexity than learning the full joint distribution.

%\revision{\info{R4:O2} To assess ECN performance under sparse workloads, we retrained it with (i) 25\% of the queries (20K) and (ii) dimensionality-aware sparsification, retaining a fraction $p^d$ of queries of dimensionality $d$. As shown in Fig.~\ref{fig:sparsewkldexp}, accuracy drops only slightly with 20K queries and degrades moderately under sparsification (notably at $p=0.9$), confirming that ECN generalizes well and remains robust even with sparse training data.}
 To evaluate ECN under sparse workloads, we retrain it using (i) 25\% of the queries (20K) and (ii) dimensionality-aware sparsification, keeping a fraction $p^d$ of queries with dimensionality $d$. As shown in Fig.~\ref{fig:sparsewkldexp}, accuracy drops only slightly with 20K queries and moderately under sparsification (especially at $p=0.9$), indicating that ECN generalizes well and remains robust even with limited training data.

\vspace{-0.5em}
\subsection{Evaluation of Join Extension}
The join extension was evaluated on 
the IMDB Job-light workload (70 queries over 6 tables)~\cite{kipf_job_light_sql}, with comparisons to two state-of-the-art baselines, ASM~\cite{kim2024asm} and PRICE~\cite{zeng2024price}. According to the results in Table~\ref{tab:joinresults}, the extended CoLSE achieves the lowest end-to-end execution time (356 s) compared with ASM (374 s) and PRICE (366 s). Inference latency was markedly lower at 1.25 ms per query, versus 9.84 ms for ASM and 20.13 ms for PRICE. Training was also faster, completing in 5 min compared to 38 min for ASM and 10 min for PRICE, while the model size remained compact (28 MB vs. 42 MB for both baselines). Overall, these results demonstrate strong efficiency and scalability without compromising accuracy. Future work will consider extensions to multi-attribute and non-equi joins.

The latest work on join cardinality estimation, LpBound~\cite{zhang2025lpbound}, computes upper bounds for cardinality. While not a direct competitor, we evaluate CoLSE with the q-error–style metric suggested in the LpBound paper. CoLSE consistently outperforms LpBound: median error drops from 5.25 to 1.62 ($\sim$3.2×), 75th percentile from 19.05 to 2.12 ($\sim$9.0×), and max from 63.1 to 10.33 ($\sim$6.1×), with lower variability (IQR: 1.39–2.12 vs. 2.63–19.05). These gaps reflect design choices: LpBound always overestimates; whereas CoLSE occasionally underestimates but achieves substantially higher accuracy and stability overall.

% \rewrite{
% \info{Can we revise this section? and save around 1/4 page}}
\subsection{Evaluation Insights}

%To holistically assess model performance, we compare CoLSE with baseline methods across multiple dimensions: accuracy, training time, inference latency, and model size. Because these metrics vary in scale and direction (i.e., higher or lower is better), we normalize all dimensions to the range [0,1] for fair visualization in spider charts. Specifically, we use min-max normalization for metrics where higher values are better (e.g., Accuracy) and reverse the scale for metrics where lower values are preferred (e.g., Inference Time, Model Size, Training Time).
To comprehensively compare CoLSE with baselines, we evaluate accuracy, training time, inference latency, and model size. Since these metrics differ in scale and optimization direction, we normalize all to [0,1] for fair spider chart visualization, applying min-max normalization for metrics where higher is better (e.g., accuracy) and inverting the scale for those where lower is better (e.g., inference and training time).

The spider charts in Fig.~\ref{fig:spider} illustrate CoLSE’s performance trade-offs and robustness across a range of scenarios. In both Fig.~\ref{fig:spider}a and Fig.~\ref{fig:spider}b, CoLSE achieves faster training, smaller model size, and lower inference latency than all baselines, while sacrificing only a small amount of accuracy. In contrast, other models typically degrade significantly in at least one of these dimensions when trying to match CoLSE's accuracy.

Under correlated data distributions, CoLSE shows only a slight dip in accuracy while maintaining resource efficiency, unlike most baseline models. When the query workload distribution shifts (Fig.~\ref{fig:spider}d), CoLSE continues to perform consistently, demonstrating robustness on par with data-driven methods—an area where query-driven models tend to falter.

%On large datasets, CoLSE retains or even improves its accuracy with only modest increases in training time and model size. By comparison, data-driven baselines incur much steeper resource costs. Taken together, these results confirm that CoLSE delivers a uniquely balanced combination of accuracy, scalability, and resilience: it adapts well to growing data volumes, handles correlation gracefully, and remains robust under dynamic workloads.
On large datasets, CoLSE maintains, or even improves, its accuracy with only modest increases in training time and model size. In contrast, data-driven baselines face significantly higher resource costs. 
Beyond aggregate metrics, CoLSE remains robust under dynamic workloads: it preserves strong performance after data updates even without retraining and, while not always best post-retraining, consistently outperforms query-driven methods and leads all models on skewed data appends.
Overall, CoLSE offers a uniquely balanced mix of accuracy, scalability, and robustness.

% Overall, CoLSE offers a uniquely balanced mix of accuracy, scalability, and robustness: it scales effectively with data volume, handles correlations well, and remains resilient under dynamic workloads.

%With data updates, CoLSE demonstrates the highest robustness among all baselines, maintaining superior performance even without retraining. While it does not outperform every method after retraining, it consistently surpasses query-driven approaches and outperforms all models in the specific case of appending skewed data.
% CoLSE shows the highest robustness to data updates, maintaining strong performance even without retraining. While it may not outperform all methods post-retraining, it consistently exceeds query-driven approaches and leads all models when handling skewed data appends.

%These findings also highlight limitations in current learned estimators. Data-driven models often excel in accuracy but suffer from high inference latency and inconsistent memory usage. Query-driven methods, while efficient at inference, rely on pre-collected workloads and lack generalizability. For instance, LW-XGB trains quickly but consumes substantial memory, while Naru offers high accuracy but is prohibitively slow to train. Furthermore, the need for extensive hyperparameter tuning reduces the practicality of these models in real-world DBMS deployments.
These findings also underscore the limitations of current learned estimators. Data-driven models often achieve high accuracy but struggle with high inference latency and unpredictable memory usage. Query-driven methods are inference-efficient but depend on pre-collected workloads and lack generalizability. For example, LW-XGB trains quickly but uses significant memory, while Naru delivers high accuracy but is extremely slow to train. Moreover, the need for extensive hyperparameter tuning limits their practicality in DBMSs.
% in real-world DBMS deployments.

%CoLSE addresses these shortcomings by combining the strengths of both paradigms. It employs a novel copula-based algorithm to model joint distributions via CDFs and refines its predictions with a lightweight error compensation network. Across all datasets, CoLSE consistently delivers competitive accuracy, sub-2ms inference latency, training times under 15 minutes, and compact model sizes under 4 MB—without requiring any hyperparameter tuning. These characteristics make CoLSE a practical and deployable solution for modern database systems.
% CoLSE bridges the gap between data- and query-driven approaches by combining their strengths. It uses a novel copula-based algorithm to model joint distributions via CDFs and enhances predictions with a lightweight error compensation network. Across all datasets, CoLSE consistently achieves competitive accuracy, sub-2ms inference latency, training times under 15 minutes, and model sizes below 4MB—without any hyperparameter tuning. These qualities make CoLSE a practical, deployable solution for modern DBMSs.
Overall, CoLSE bridges data- and query-driven approaches by combining their strengths, which helps it to deliver competitive accuracy with sub-2 ms inference, $\le$15 min training, and $\le$4 MB model size—without hyperparameter tuning. Crucially, CoLSE scales cleanly to large datasets (tens of GB) with only modest resource growth and remains robust under correlations, workload shifts, and skewed data appends, often without retraining, making it a practical deployable choice for modern DBMSs.

\vspace{-0.5em}
\section{Conclusion}

CoLSE presents a novel hybrid architecture for single-table cardinality estimation, integrating data-driven modeling with query-driven correction. Its core innovation lies in approximating joint distributions through CDFs and pairwise D-vine copulas, avoiding the complexity of high-dimensional PDFs while preserving key attribute dependencies.

This is complemented by an error compensation network that adjusts estimates based on query workload signals, enhancing accuracy without the need for hyperparameter tuning. Together, these components allow CoLSE to achieve a strong balance of accuracy, efficiency, and scalability, outperforming existing methods in diverse scenarios including skewed, correlated, and shifting workloads. 
%For future work, we plan to explore online or incremental learning to reduce retraining overhead during data updates. 

% Additionally, we aim to extend the copula-based framework to support join queries, further broadening CoLSE’s applicability.

\section*{Acknowledgments}

We gratefully acknowledge support from the Australian Research Council Discovery Early Career Researcher Award DE230100366, and Google Foundational Science 2025 fund.

\section{AI-Generated Content Acknowledgement}
The authors used ChatGPT to assist with English language editing, improving the grammar, and clarity of text originally written by the authors. Furthermore, ChatGPT was used during code development to assist with basic syntax and implementation details. No experimental results or novel research content were generated solely by AI tools. The authors take full responsibility for the accuracy, originality, and integrity of all content presented in this paper.

\bibliographystyle{IEEEtran}
\balance
\bibliography{IEEEabrv}

@String{Springer = "Springer-Verlag" }

@Article{Wang2021face,
  title={FACE: A normalizing flow based cardinality estimator},
  author={Wang, Jiayi and Chai, Chengliang and Liu, Jiabin and Li, Guoliang},
  journal={Proceedings of the VLDB Endowment},
  volume={15},
  number={1},
  pages={72--84},
  year={2021},
  publisher={VLDB Endowment}
}

@article{lan2021survey,
  title={A survey on advancing the dbms query optimizer: Cardinality estimation, cost model, and plan enumeration},
  author={Lan, Hai and Bao, Zhifeng and Peng, Yuwei},
  journal={Data Science and Engineering},
  volume={6},
  pages={86--101},
  year={2021},
  publisher={Springer}
}

@article{poosala1996improved,
  title={Improved histograms for selectivity estimation of range predicates},
  author={Poosala, Viswanath and Haas, Peter J and Ioannidis, Yannis E and Shekita, Eugene J},
  journal={ACM Sigmod Record},
  volume={25},
  number={2},
  pages={294--305},
  year={1996},
  publisher={ACM New York, NY, USA}
}

@article{leis2015good,
  title={How good are query optimizers, really?},
  author={Leis, Viktor and Gubichev, Andrey and Mirchev, Atanas and Boncz, Peter and Kemper, Alfons and Neumann, Thomas},
  journal={Proceedings of the VLDB Endowment},
  volume={9},
  number={3},
  pages={204--215},
  year={2015},
  publisher={VLDB Endowment}
}

@inproceedings{park2020quicksel,
  title={Quicksel: Quick selectivity learning with mixture models},
  author={Park, Yongjoo and Zhong, Shucheng and Mozafari, Barzan},
  booktitle={Proceedings of the 2020 ACM SIGMOD International Conference on Management of Data},
  pages={1017--1033},
  year={2020}
}

@article{dutt2019selectivity,
  title={Selectivity estimation for range predicates using lightweight models},
  author={Dutt, Anshuman and Wang, Chi and Nazi, Azade and Kandula, Srikanth and Narasayya, Vivek and Chaudhuri, Surajit},
  journal={Proceedings of the VLDB Endowment},
  volume={12},
  number={9},
  pages={1044--1057},
  year={2019},
  publisher={VLDB Endowment}
}

@inproceedings{hasan2020deep,
  title={Deep learning models for selectivity estimation of multi-attribute queries},
  author={Hasan, Shohedul and Thirumuruganathan, Saravanan and Augustine, Jees and Koudas, Nick and Das, Gautam},
  booktitle={Proceedings of the 2020 ACM SIGMOD International Conference on Management of Data},
  pages={1035--1050},
  year={2020}
}

@article{kipf2018learned,
  title={Learned cardinalities: Estimating correlated joins with deep learning},
  author={Kipf, Andreas and Kipf, Thomas and Radke, Bernhard and Leis, Viktor and Boncz, Peter and Kemper, Alfons},
  journal={arXiv preprint arXiv:1809.00677},
  year={2018}
}

@inproceedings{heimel2015self,
  title={Self-tuning, gpu-accelerated kernel density models for multidimensional selectivity estimation},
  author={Heimel, Max and Kiefer, Martin and Markl, Volker},
  booktitle={Proceedings of the 2015 ACM SIGMOD International Conference on Management of Data},
  pages={1477--1492},
  year={2015}
}

@article{yang2019deep,
  title={Deep unsupervised cardinality estimation},
  author={Yang, Zongheng and Liang, Eric and Kamsetty, Amog and Wu, Chenggang and Duan, Yan and Chen, Xi and Abbeel, Pieter and Hellerstein, Joseph M and Krishnan, Sanjay and Stoica, Ion},
  journal={arXiv preprint arXiv:1905.04278},
  year={2019}
}

@article{hilprecht2019deepdb,
  title={Deepdb: Learn from data, not from queries!},
  author={Hilprecht, Benjamin and Schmidt, Andreas and Kulessa, Moritz and Molina, Alejandro and Kersting, Kristian and Binnig, Carsten},
  journal={arXiv preprint arXiv:1909.00607},
  year={2019}
}

@article{wang2020we,
  title={Are we ready for learned cardinality estimation?},
  author={Wang, Xiaoying and Qu, Changbo and Wu, Weiyuan and Wang, Jiannan and Zhou, Qingqing},
  journal={arXiv preprint arXiv:2012.06743},
  year={2020}
}

@article{aas2009pair,
  title={Pair-copula constructions of multiple dependence},
  author={Aas, Kjersti and Czado, Claudia and Frigessi, Arnoldo and Bakken, Henrik},
  journal={Insurance: Mathematics and economics},
  volume={44},
  number={2},
  pages={182--198},
  year={2009},
  publisher={Elsevier}
}

@inproceedings{germain2015made,
  title={Made: Masked autoencoder for distribution estimation},
  author={Germain, Mathieu and Gregor, Karol and Murray, Iain and Larochelle, Hugo},
  booktitle={International conference on machine learning},
  pages={881--889},
  year={2015},
  organization={PMLR}
}

@article{han2021transformer,
  title={Transformer in transformer},
  author={Han, Kai and Xiao, An and Wu, Enhua and Guo, Jianyuan and Xu, Chunjing and Wang, Yunhe},
  journal={Advances in neural information processing systems},
  volume={34},
  pages={15908--15919},
  year={2021}
}

@article{han2021cardinality,
  title={Cardinality estimation in dbms: A comprehensive benchmark evaluation},
  author={Han, Yuxing and Wu, Ziniu and Wu, Peizhi and Zhu, Rong and Yang, Jingyi and Tan, Liang Wei and Zeng, Kai and Cong, Gao and Qin, Yanzhao and Pfadler, Andreas and others},
  journal={arXiv preprint arXiv:2109.05877},
  year={2021}
}

@INPROCEEDINGS{9094107,
  author={Negi, Parimarjan and Marcus, Ryan and Mao, Hongzi and Tatbul, Nesime and Kraska, Tim and Alizadeh, Mohammad},
  booktitle={2020 IEEE 36th International Conference on Data Engineering Workshops (ICDEW)}, 
  title={Cost-Guided Cardinality Estimation: Focus Where it Matters}, 
  year={2020},
  volume={},
  number={},
  pages={154-157},
  doi={10.1109/ICDEW49219.2020.00034}
}

@article{negi2021flow,
  title={Flow-loss: Learning cardinality estimates that matter},
  author={Negi, Parimarjan and Marcus, Ryan and Kipf, Andreas and Mao, Hongzi and Tatbul, Nesime and Kraska, Tim and Alizadeh, Mohammad},
  journal={arXiv preprint arXiv:2101.04964},
  year={2021}
}

@article{ling2020deep,
  title={Deep archimedean copulas},
  author={Ling, Chun Kai and Fang, Fei and Kolter, J Zico},
  journal={Advances in Neural Information Processing Systems},
  volume={33},
  pages={1535--1545},
  year={2020}
}

@article{cormode2011synopses,
  title={Synopses for massive data: Samples, histograms, wavelets, sketches},
  author={Cormode, Graham and Garofalakis, Minos and Haas, Peter J and Jermaine, Chris and others},
  journal={Foundations and Trends{\textregistered} in Databases},
  volume={4},
  number={1--3},
  pages={1--294},
  year={2011},
  publisher={Now Publishers, Inc.}
}

@misc{Wikipedia,
  author = {Wikipedia contributors},
  title = {Copula (statistics)},
  year = {2025},
  url = {\url{https://en.wikipedia.org/wiki/Copula_(statistics)}},
  note = {Accessed: 2025-05-03}
}

@article{gorecki2016approach,
  title={An approach to structure determination and estimation of hierarchical Archimedean copulas and its application to Bayesian classification},
  author={G{\'o}recki, Jan and Hofert, Marius and Hole{\v{n}}a, Martin},
  journal={Journal of Intelligent Information Systems},
  volume={46},
  number={1},
  pages={21--59},
  year={2016},
  publisher={Springer}
}

@incollection{sane2013inclusion,
  title={The inclusion-exclusion principle},
  author={Sane, Sharad S},
  booktitle={Combinatorial Techniques},
  pages={57--79},
  year={2013},
  publisher={Springer}
}

@book{poosala1997histogram,
  title={Histogram-based estimation techniques in database systems},
  author={Poosala, Viswanath},
  year={1997},
  publisher={The University of Wisconsin-Madison}
}

@inproceedings{poosala1997selectivity,
  title={Selectivity estimation without the attribute value independence assumption},
  author={Poosala, Viswanath and Ioannidis, Yannis E},
  booktitle={VLDB},
  volume={97},
  pages={486--495},
  year={1997}
}

@inproceedings{bruno2001stholes,
  title={STHoles: A multidimensional workload-aware histogram},
  author={Bruno, Nicolas and Chaudhuri, Surajit and Gravano, Luis},
  booktitle={Proceedings of the 2001 ACM SIGMOD international conference on Management of data},
  pages={211--222},
  year={2001}
}

@article{czado2022vine,
  title={Vine copula based modeling},
  author={Czado, Claudia and Nagler, Thomas},
  journal={Annual Review of Statistics and Its Application},
  volume={9},
  number={1},
  pages={453--477},
  year={2022},
  publisher={Annual Reviews}
}

@misc{Suzuki2024CostEstimation,
  author       = {Hironobu Suzuki},
  title        = {3.2. Cost Estimation in Single-Table Query},
  howpublished = {\url{https://www.interdb.jp/pg/pgsql03/02.html}},
  year         = {2024},
  note         = {Accessed: 2025-05-03}
}

@misc{fzirak2025tpchskew,
  author       = {Farhad Zirak},
  title        = {TPCH-Skew for Linux},
  howpublished = {\url{https://github.com/fzirak/tpch-skew-linux}},
  year         = {2025},
  note         = {Accessed: 2025-05-03}
}

@manual{PostgreSQL13Documentation,
  title        = {PostgreSQL 13.20 Documentation},
  author       = {{The PostgreSQL Global Development Group}},
  organization = {PostgreSQL Global Development Group},
  year         = {2025},
  url          = {https://www.postgresql.org/docs/13/index.html},
  note         = {Accessed: 2025-05-03}
}

@article{genest1993statistical,
  title={Statistical inference procedures for bivariate Archimedean copulas},
  author={Genest, Christian and Rivest, Louis-Paul},
  journal={Journal of the American statistical Association},
  volume={88},
  number={423},
  pages={1034--1043},
  year={1993},
  publisher={Taylor \& Francis}
}

@misc{wikipedia_probability_integral_transform,
  title        = {Probability integral transform},
  author       = {{Wikipedia contributors}},
  year         = {2025},
  month        = {May},
  url          = {\url{https://en.wikipedia.org/wiki/Probability_integral_transform}},
  note         = {Accessed on 2025-05-03},
}

@inproceedings{wu2021unified,
  title={A unified deep model of learning from both data and queries for cardinality estimation},
  author={Wu, Peizhi and Cong, Gao},
  booktitle={Proceedings of the 2021 International Conference on Management of Data},
  pages={2009--2022},
  year={2021}
}

@inproceedings{lohman2014query,
  title={Is query optimization a “solved” problem},
  author={Lohman, Guy},
  booktitle={Proc. Workshop on Database Query Optimization},
  volume={13},
  pages={10},
  year={2014},
  organization={Oregon Graduate Center Comp. Sci. Tech. Rep}
}

@article{lee2023analyzing,
  title={Analyzing the impact of cardinality estimation on execution plans in microsoft SQL server},
  author={Lee, Kukjin and Dutt, Anshuman and Narasayya, Vivek and Chaudhuri, Surajit},
  journal={Proceedings of the VLDB Endowment},
  volume={16},
  number={11},
  pages={2871--2883},
  year={2023},
  publisher={VLDB Endowment}
}

@article{wu2023factorjoin,
  title={FactorJoin: a new cardinality estimation framework for join queries},
  author={Wu, Ziniu and Negi, Parimarjan and Alizadeh, Mohammad and Kraska, Tim and Madden, Samuel},
  journal={Proceedings of the ACM on Management of Data},
  volume={1},
  number={1},
  pages={1--27},
  year={2023},
  publisher={ACM New York, NY, USA}
}

@article{kim2024asm,
  title={Asm: Harmonizing autoregressive model, sampling, and multi-dimensional statistics merging for cardinality estimation},
  author={Kim, Kyoungmin and Lee, Sangoh and Kim, Injung and Han, Wook-Shin},
  journal={Proceedings of the ACM on Management of Data},
  volume={2},
  number={1},
  pages={1--27},
  year={2024},
  publisher={ACM New York, NY, USA}
}

@article{zhu2021glue,
  title={Glue: Adaptively Merging Single Table Cardinality to Estimate Join Query Size},
  author={Zhu, Rong and Zeng, Tianjing and Pfadler, Andreas and Chen, Wei and Ding, Bolin and Zhou, Jingren},
  journal={arXiv preprint arXiv:2112.03458},
  year={2021}
}

@article{zhu2020flat,
  title={FLAT: fast, lightweight and accurate method for cardinality estimation},
  author={Zhu, Rong and Wu, Ziniu and Han, Yuxing and Zeng, Kai and Pfadler, Andreas and Qian, Zhengping and Zhou, Jingren and Cui, Bin},
  journal={arXiv preprint arXiv:2011.09022},
  year={2020}
}

@article{lin2023cardinality,
  title={Cardinality estimation with smoothing autoregressive models},
  author={Lin, Yuming and Xu, Zejun and Zhang, Yinghao and Li, You and Zhang, Jingwei},
  journal={World Wide Web},
  volume={26},
  number={5},
  pages={3441--3461},
  year={2023},
  publisher={Springer}
}

@inproceedings{kim2022learned,
  title={Learned cardinality estimation: An in-depth study},
  author={Kim, Kyoungmin and Jung, Jisung and Seo, In and Han, Wook-Shin and Choi, Kangwoo and Chong, Jaehyok},
  booktitle={Proceedings of the 2022 international conference on management of data},
  pages={1214--1227},
  year={2022}
}

@inproceedings{getoor2001selectivity,
  title={Selectivity estimation using probabilistic models},
  author={Getoor, Lise and Taskar, Benjamin and Koller, Daphne},
  booktitle={Proceedings of the 2001 ACM SIGMOD international conference on Management of data},
  pages={461--472},
  year={2001}
}

@article{tzoumas2011lightweight,
  title={Lightweight graphical models for selectivity estimation without independence assumptions},
  author={Tzoumas, Kostas and Deshpande, Amol and Jensen, Christian S},
  journal={Proceedings of the VLDB Endowment},
  volume={4},
  number={11},
  pages={852--863},
  year={2011},
  publisher={VLDB Endowment}
}

@article{scanagatta2019survey,
  title={A survey on Bayesian network structure learning from data},
  author={Scanagatta, Mauro and Salmer{\'o}n, Antonio and Stella, Fabio},
  journal={Progress in Artificial Intelligence},
  volume={8},
  number={4},
  pages={425--439},
  year={2019},
  publisher={Springer}
}

@article{wu2001applying,
  title={Applying the golden rule of sampling for query estimation},
  author={Wu, Yi-Leh and Agrawal, Divyakant and El Abbadi, Amr},
  journal={ACM SIGMOD Record},
  volume={30},
  number={2},
  pages={449--460},
  year={2001},
  publisher={ACM New York, NY, USA}
}

@inproceedings{muralikrishna1988equi,
  title={Equi-depth multidimensional histograms},
  author={Muralikrishna, M and DeWitt, David J},
  booktitle={Proceedings of the 1988 ACM SIGMOD international conference on Management of data},
  pages={28--36},
  year={1988}
}

@article{aboulnaga1999self,
  title={Self-tuning histograms: Building histograms without looking at data},
  author={Aboulnaga, Ashraf and Chaudhuri, Surajit},
  journal={ACM SIGMOD Record},
  volume={28},
  number={2},
  pages={181--192},
  year={1999},
  publisher={ACM New York, NY, USA}
}

@article{brechmann2013modeling,
  title={Modeling dependence with C-and D-vine copulas: the R package CDVine},
  author={Brechmann, Eike Christian and Schepsmeier, Ulf},
  journal={Journal of statistical software},
  volume={52},
  pages={1--27},
  year={2013}
}

@online{postgresqlrowestimationexamples,
  author       = {{PostgreSQL Global Development Group}},
  title        = {Row Estimation Examples},
  year         = {2024},
  organization = {PostgreSQL},
  url          = {https://www.postgresql.org/docs/current/row-estimation-examples.html},
  urldate      = {2025-06-17}
}

@article{alece,
author = {Li, Pengfei and Wei, Wenqing and Zhu, Rong and Ding, Bolin and Zhou, Jingren and Lu, Hua},
title = {ALECE: An Attention-based Learned Cardinality Estimator for SPJ Queries on Dynamic Workloads},
year = {2023},
issue_date = {October 2023},
publisher = {VLDB Endowment},
volume = {17},
number = {2},
issn = {2150-8097},
url = {https://doi.org/10.14778/3626292.3626302},
doi = {10.14778/3626292.3626302},
journal = {Proc. VLDB Endow.},
pages = {197–210},
numpages = {14}
}

@article{dunn1996randomized,
  title={Randomized quantile residuals},
  author={Dunn, Peter K and Smyth, Gordon K},
  journal={Journal of Computational and graphical statistics},
  volume={5},
  number={3},
  pages={236--244},
  year={1996},
  publisher={Taylor \& Francis}
}

@incollection{Mizuno2022SIS,
  author    = {Mizuno, T. and Deutsch, C.},
  title     = {Sequential Indicator Simulation (SIS)},
  booktitle = {Geostatistics Lessons},
  editor    = {Deutsch, J.L.},
  year      = {2022},
  publisher = {Centre for Computational Geostatistics},
  url       = {http://www.geostatisticslessons.com/lessons/sequentialindicatorsim}
}

@misc{Categorical_Distribution_2025,
  title        = {Categorical distribution — Probability Distribution Explorer},
  author       = {Bois, Justin},
  howpublished = {\url{https://distribution-explorer.github.io/discrete/categorical.html?utm_source=chatgpt.com}},
  note         = {Last updated August 09, 2025},
  year         = {2025}
}

@inproceedings{marcus2021bao,
  title={Bao: Making learned query optimization practical},
  author={Marcus, Ryan and Negi, Parimarjan and Mao, Hongzi and Tatbul, Nesime and Alizadeh, Mohammad and Kraska, Tim},
  booktitle={Proceedings of the 2021 International Conference on Management of Data},
  pages={1275--1288},
  year={2021}
}

@article{woltmann2023fastgres,
  title={Fastgres: Making learned query optimizer hinting effective},
  author={Woltmann, Lucas and Thiessat, Jerome and Hartmann, Claudio and Habich, Dirk and Lehner, Wolfgang},
  journal={Proceedings of the VLDB Endowment},
  volume={16},
  number={11},
  pages={3310--3322},
  year={2023},
  publisher={VLDB Endowment}
}

@article{xu2023coool,
  title={Coool: A learning-to-rank approach for sql hint recommendations},
  author={Xu, Xianghong and Zhao, Zhibing and Zhang, Tieying and Kang, Rong and Sun, Luming and Chen, Jianjun},
  journal={arXiv preprint arXiv:2304.04407},
  year={2023}
}

@article{zhu2023lero,
  title={Lero: A learning-to-rank query optimizer},
  author={Zhu, Rong and Chen, Wei and Ding, Bolin and Chen, Xingguang and Pfadler, Andreas and Wu, Ziniu and Zhou, Jingren},
  journal={arXiv preprint arXiv:2302.06873},
  year={2023}
}

@article{zeng2024price,
  title={PRICE: a pretrained model for cross-database cardinality estimation},
  author={Zeng, Tianjing and Lan, Junwei and Ma, Jiahong and Wei, Wenqing and Zhu, Rong and Li, Pengfei and Ding, Bolin and Lian, Defu and Wei, Zhewei and Zhou, Jingren},
  journal={arXiv preprint arXiv:2406.01027},
  year={2024}
}

@online{HudsonThamesVineCopulaIntro,
  title   = {A Practical Introduction to Vine Copula},
  author  = {{Hudson \& Thames Quantitative Research}},
  year    = {2024},
  note    = {ArbitrageLab documentation, version 1.0.0},
  url     = {https://hudson-and-thames-arbitragelab.readthedocs-hosted.com/en/latest/copula_approach/vine_copula_intro.html},
  urldate = {2025-09-20}
}

@online{Wicklin2021CopulasIntro,
  author       = {Wicklin, Rick},
  title        = {An introduction to simulating correlated data by using copulas},
  date         = {2021-07-05},
  year         = {2021},
  howpublished = {\emph{The DO Loop} (SAS Blogs), SAS Institute Inc.},
  url          = {https://blogs.sas.com/content/iml/2021/07/05/introduction-copulas.html},
  urldate      = {2025-09-20}
}

@article{CzadoNagler2022VineCopula,
  author  = {Czado, Claudia and Nagler, Thomas},
  title   = {Vine Copula Based Modeling},
  journal = {Annual Review of Statistics and Its Application},
  year    = {2022},
  volume  = {9},
  pages   = {453--477},
  doi     = {10.1146/annurev-statistics-040220-101153},
  url     = {https://tnagler.github.io/vine-arisa.pdf},
  note    = {First published as a Review in Advance on November 2, 2021}
}

@online{kipf_job_light_sql,
  author    = {A. Kipf},
  title     = {JOB-light workload},
  year      = {},
  url       = {https://github.com/andreaskipf/learnedcardinalities/blob/master/workloads/job-light.sql},
  urldate   = {2025-09-21}
}

@article{zhang2025lpbound,
  title={LpBound: Pessimistic Cardinality Estimation Using {$\ell$}p-Norms of Degree Sequences},
  author={Zhang, Haozhe and Mayer, Christoph and Abo Khamis, Mahmoud and Olteanu, Dan and Suciu, Dan},
  journal={Proceedings of the ACM on Management of Data},
  volume={3},
  number={3},
  pages={1--27},
  year={2025},
  publisher={ACM New York, NY, USA}
}

\end{document}